\documentclass[apj]{emulateapj}

\shorttitle{HOT AND COLD ISM IN NGC\,2563}
\shortauthors{RASMUSSEN ET AL.}

\begin{document}

\title{HOT AND COLD GALACTIC GAS IN THE NGC\,2563 GALAXY GROUP}

\author{Jesper Rasmussen,\altaffilmark{1} Xue-Ning
  Bai,\altaffilmark{2} John S.~Mulchaey,\altaffilmark{3} J.~H.~van
  Gorkom,\altaffilmark{4} Tesla E.\ Jeltema,\altaffilmark{5} Ann
  I.\ Zabludoff,\altaffilmark{6} Eric Wilcots,\altaffilmark{7} 
  Paul Martini,\altaffilmark{8} Duane Lee\altaffilmark{4} and Timothy
  P.\ Roberts\altaffilmark{9}}

\altaffiltext{1}{Dark Cosmology Centre, Niels Bohr Institute,
  University of Copenhagen, Juliane Maries Vej 30, DK-2100 Copenhagen,
  Denmark; jr@dark-cosmology.dk}

\altaffiltext{2}{Department of Astrophysical Sciences, Peyton Hall,
  Princeton University, NJ 08544} 

\altaffiltext{3}{Carnegie Observatories, 813 Santa Barbara Street,
  Pasadena, CA 91101, USA}

\altaffiltext{4}{Department of Astronomy, Columbia University, Mail
  Code 5246, 550 West 120th Street, New York, NY 10027, USA}

\altaffiltext{5}{UCO/Lick Observatories, 1156 High Street, Santa Cruz,
  CA 95064, USA}

\altaffiltext{6}{Steward Observatory, University of Arizona, 933 North
  Cherry Avenue, Tucson, AZ 85721, USA}

\altaffiltext{7}{Department of Astronomy, University of
  Wisconsin-Madison, 475 N. Charter St., Madison, WI 53706, USA}

\altaffiltext{8}{Department of Astronomy, 4055 McPherson Laboratory,
  Ohio State University, 140 West 18th Avenue, Columbus, OH, USA}

\altaffiltext{9}{Department of Physics, Durham University, South Road,
  Durham DH1 3LE, UK}

\begin{abstract}

The role of environmentally induced gas stripping in driving galaxy
evolution in groups remains poorly understood. Here we present
extensive {\em Chandra} and Very Large Array mosaic observations of
the hot and cold interstellar medium within the members of the nearby,
X-ray bright NGC\,2563 group, a prime target for studies of the role
of gas stripping and interactions in relatively small host halos. Our
observations cover nearly all group members within a projected radius
of 1.15~Mpc ($\sim 1.4 R_{\rm vir}$) of the group center, down to a
limiting X-ray luminosity and H{\sc i} mass of $3\times
10^{39}$~erg~s$^{-1}$ and $2\times 10^8 M_\odot$, respectively. The
X-ray data are consistent with efficient ram pressure stripping of the
hot gas halos of early-type galaxies near the group core, but no X-ray
tails are seen and the limited statistics preclude strong
conclusions. The H{\sc i} results suggest moderate H{\sc i} mass loss
from the group members when compared to similar field galaxies. Six of
the 20 H{\sc i}--detected group members show H{\sc i} evidence of
ongoing interactions with other galaxies or with the intragroup
medium. Suggestive evidence is further seen for galaxies with close
neighbors in position--velocity space to show relatively low H{\sc i}
content, consistent with tidal removal of H{\sc i}. The results thus
indicate removal of both hot and cold gas from the group members via a
combination of ram pressure stripping and tidal interactions. We also
find that 16 of the 20 H{\sc i} detections occur on one side of the
group, reflecting an unusual morphological segregation whose origin
remains unclear.

\end{abstract}

\keywords{galaxies: halos --- galaxies: ISM --- galaxies: clusters:
  general --- X-rays: galaxies --- radio lines: ISM}

\section{Introduction}

Understanding the physical processes involved in the evolution of
galaxies is a key goal of extragalactic astronomy. Although stellar
(or halo) mass is emerging as playing a fundamental role for galaxy
evolution, environmental influences may also have an impact
\citep{kauf03a,kauf04,bald06,cucc10,peng10}. This is particularly
relevant in dense environments, in which galaxies may experience a
wide range of externally driven processes such as mergers, tidal
interactions, and gas stripping due to interactions with ambient
gas. All of these processes may act to remove a substantial fraction
of the interstellar medium (ISM), or rapidly consume or eject it
through interaction--induced star formation and nuclear activity,
eventually leading the galaxy to transition from blue and star-forming
to red and quiescent \citep{miho96,quil00,verd01,chun07,kawa08}

The relative importance of these various processes should itself be a
function of environment, with mergers and tidal interactions
dominating in small galaxy groups (e.g., \citealt{barn89,mamo07}), and
with galaxy harassment, ram pressure stripping, and starvation (the
cut-off of the supply of cold gas for star formation from warm/hot gas
in the halo) becoming more efficient in massive clusters exhibiting
high galaxy velocities \citep{moor99,quil00}. {\em Chandra}
observations do show that a lower fraction of early-type galaxies in
clusters contain hot halos than their counterparts in groups
\citep{jelt08}, consistent with expectations of the hot halo stripping
efficiency being higher in more massive systems. Evidence for ram
pressure stripping of the {\em cold} ISM component in rich clusters is
also well-established through many H{\sc i} studies
\citep{cham80,caya94,schr01,sola01,chun07,levy07,chun09,cort11},
suggesting that interactions between galaxies and the intracluster
medium may generally be an important route to removing galactic gas in
the most massive systems.

The situation is less clear in smaller groups, despite these
representing much more typical galaxy environments. Naively,
galaxy--galaxy interactions and mergers should be relatively more
important, given the lower galaxy velocities and intergalactic medium
densities in groups. Indeed, the results of \citet{jelt08} and
\citet{mulc10} show that hot X-ray halos are retained around the
majority of $L_K>L^\ast$ early-types in the central regions of groups,
and these halos are not strongly X-ray underluminous compared to those
of field galaxies. Hence, removal of hot halo gas by ram pressure must
be very modest in groups, at least for massive galaxies. Nevertheless,
numerical simulations suggest that starvation may still occur for
moderate-luminosity galaxies on their first passage through even
fairly small groups \citep{kawa08}. Viscous stripping of galactic gas
through Kelvin--Helmholtz instabilities, the efficiency of which
depends only linearly on galaxy velocity \citep{nuls82}, could also
play a role in groups even when brute-force ram pressure is
unimportant \citep{rasm08}.

For the cold gas, a number of group galaxies show evidence for
extended H{\sc i} that has been stripped from their host (e.g.,
\citealt{verd01,kant05,kilb06}), and many are deficient in H{\sc i}
compared to similar field galaxies \citep{verd01,seng07,kilb09}. Some
of these objects represent strong candidates for ram pressure
stripping of both cold and hot ISM
\citep{bure02,rasm06,seng07,mcco07,bail07}. However, the H{\sc i}
properties of spirals in these environments can often be equally well
explained by tidal encounters \citep{kern08,rasm08,kilb09}, and so it
is still not fully clear which process, if any, dominates the gas
removal from typical group galaxies.

Quantifying the importance of the various mechanisms acting on group
galaxies requires detailed observations of the ISM in these galaxies,
both within individual groups and across systems displaying a range of
global properties. To identify signatures of ongoing gas removal and
how these may depend on local group environment, the ISM must be
probed on spatial scales of individual galaxies across the full
system, with a sensitivity extending down to moderate-luminosity
galaxies. Exploring the role of hot gas stripping, whether induced by
ram pressure or galaxy--galaxy interactions, is currently largely
limited to early-type galaxies or starbursting spirals, however, since
only these galaxy types generally contain X-ray detectable halo gas
\citep{rasm09}. In more quiescent late-type spirals, gas stripping can
be far more efficiently explored using H{\sc i} data. Combining X-ray
and H{\sc i} data therefore allows one to study the ISM over the full
galaxy morphological range.

Here we combine X-ray mosaicing observations from {\em Chandra} with
analogous H{\sc i} observations from the Very Large
Array\footnote{The Very Large Array is operated by the National
  Radio Astronomy Observatory, which is a facility of the National
  Science Foundation (NSF), operated under cooperative agreement by
  Associated Universities, Inc.} (VLA) to study the hot and cold ISM
in the nearby, X-ray bright NGC\,2563 group.  Our goal is to search
for evidence of ongoing or recent gas stripping (we here use this term
to mean any externally induced removal of gas from galaxies, including
via tidal stripping), explore possible mechanisms involved, and
understand their environmental dependence. While X-ray
\citep{fabb92,mulc98,osmo04,gast07} and optical \citep{zabl98a,zabl00}
observations of this group already exist, a unique aspect of the
present study is that both our X-ray and H{\sc i} observations cover
the entire group (out to a projected radius of $R=1.15$~Mpc, well
beyond the estimated virial radius), while still benefiting from the
sub-kpc spatial resolution of {\em Chandra} at the target
redshift. Combined with the sensitivity of the VLA to low--surface
brightness diffuse H{\sc i} emission (with a synthesized beam width of
$\approx 45''$ in the most compact ``D-array'' conguration employed
here), this enables a detailed investigation of the spatial variation
in global ISM properties across the full group environment.

We describe the overall properties of the group in
Section~\ref{sec,group}. The X-ray and H{\sc i} analyses are detailed
in Section~\ref{sec,analysis}, and results are presented in
Section~\ref{sec,results}. Section~\ref{sec,discuss} considers the
evidence for recent and past ISM stripping in the group, and our
results and conclusions are summarized in Section~\ref{sec,summary}.
We assume $H_0=70$~km~s$^{-1}$~Mpc$^{-1}$, $\Omega_{m}=0.27$, and
$\Omega_{\Lambda}=0.73$. The target redshift of $z=0.0157$ then
corresponds to a luminosity distance of 68.1~Mpc, and $1'$ on the sky
to a projected distance of 19.2~kpc. Uncertainties are quoted at the
$1\sigma$ level unless otherwise specified.

\vspace{4mm}

\section{Group Membership and Properties}\label{sec,group}

Extensive optical spectroscopy of the NGC\,2563 field exists from the
studies of \citet{zabl98a,zabl00}. The field is also covered by the
Sloan Digital Sky Survey (SDSS), providing images and magnitudes for
all group members, and flux-calibrated spectra for most of them. From
the measured redshifts, we determined group membership using the
ROSTAT package \citep{beer90}, considering all galaxies with known
recessional velocities within $\pm 3000$~km~s$^{-1}$ of the group
mean.  We then calculated the biweight estimators of velocity mean and
dispersion.  Objects with velocities beyond $\pm 3\sigma$$_{\rm biwt}$
from the mean were discarded, and the process repeated iteratively
until convergence. This technique identified 64 group members within a
projected radius of $R=60'$ ($R = 1.15$~Mpc) from the group center,
including two newly discovered members from our H{\sc i} observations
(see Section~\ref{sec,HI}). The resulting mean redshift and radial
velocity dispersion of the group based on these 64 galaxies are
$z=0.0157 \pm 0.0001$ and $\sigma_{\rm biwt} =
364^{+36}_{-33}$~km~s$^{-1}$, respectively. Based on objects
classified as galaxies in the SDSS survey, the group membership is
100\% (98\%) spectroscopically complete down to $M_r=-18$ ($-17$)
within $60'$ of the group center.

Group members were morphologically classified independently by JSM and
AIZ using SDSS $r$-band images. The two authors' classifications
agreed within one Hubble type in all cases, and our adopted
morphologies represent the average of the two results. We use SDSS
$g$-band magnitudes as a measure of the blue luminosity, and, for
comparison to previous studies, $K_s$ magnitudes from the 2-Micron All
Sky Survey (2MASS) as a rough estimate of galaxy stellar mass. The
latter magnitudes are available for most of the group members, except
for a few galaxies in close pairs. In these cases, the $K_s$ magnitude
was estimated from the SDSS $r$-band magnitude using $r$--$K_s=2.90$
for early-type galaxies, and $r$--$K_s=2.30$ for late-types, based on
the average values for the group members with both magnitudes
available. The scatter in each case is $\sim$\,0.20. As none of the
galaxies without $K_s$ magnitudes is detected in X-rays or H{\sc i},
adopting these relationships has minimal impact on our
conclusions. Magnitudes were converted to solar luminosities assuming
$g_{\bigodot}=5.12$ and $K_{\bigodot}=3.39$ after correcting for
Galactic extinction. We further adopt $L_g^\ast = 10^{10.17} L_\odot$
\citep{mont09} and $L_K^\ast =10^{11.08} L_\odot$ \citep{sun07}.

While a detailed X-ray mass analysis of the group is beyond the scope
of this paper, having a rough estimate of the group virial radius
$R_{\rm vir}$ is instructive for the discussion to follow. This can be
obtained from the hot gas temperature of $kT \approx 1.1$~keV derived
by \citet{mulc03}, combined with the scaling relation of
\citet{fino07}. This would suggest $R_{500} \sim 420$~kpc, and hence
$R_{\rm vir}\approx R_{100}\approx 2R_{500} \approx 850$~kpc in the
adopted cosmology, assuming a Navarro--Frenk--White potential
\citep{nava97} with a typical concentration parameter of
$c=5$--10. Here $R_\Delta$ is the radius enclosing a mean density of
$\Delta$ times the critical value.

\vspace{3mm}

\section{Observations and Analysis}\label{sec,analysis}

\vspace{1mm}

\subsection{{\em Chandra} Imaging and Spectroscopy}\label{sec,spec}

The distribution of group members in NGC\,2563 allows most of them to
be observed with 14 ACIS-I pointings: a $3\times3$ central mosaic to
cover the inner $45'\times45'$ of the group and five outer
pointings. A total of 54 of the 64 confirmed group members were
observed by {\em Chandra}. The majority of the remaining galaxies are
optically faint and therefore not likely to be strong X-ray
sources. The 14 pointings were done between June 2007 and March 2008
for 30~ks each, except for the central pointing which was observed for
50~ks to achieve a similar contrast for galaxies in the X-ray bright
central region. All observations were conducted in VFAINT telemetry
mode. The level~1 event files were processed with {\sc
  ciao}\footnote{http://cxc.harvard.edu/ciao/} v4.0 and
CALDB\footnote{http://cxc.harvard.edu/caldb/} v3.4.3, filtering out
bad pixels and applying the latest gain files.  Events with {\em ASCA}
grades\footnote{see
  http://cxc.harvard.edu/ciao/ahelp/acis\_process\_events.html} 1, 5,
and 7, and a non-zero status flag were discarded.  Finally, background
flares were screened for using the routine 'lc\_clean.sl' in {\sc
  ciao}, excluding time periods with background count rates greater
than $20\%$ of the quiescent rate. Flaring was not a problem for any
of our observations.

Source detection was performed using the ``Mexican Hat'' wavelet
source detection algorithm 'wavdetect' in {\sc ciao}, with scales of
1, 2, 4, 8, 16 and 32 pixels, using both soft (0.3--2.0~keV), hard
(2.0--7.0~keV), and ``full'' band (0.3--7~keV) event lists. The
detection threshold was set to limit the number of false detections to
$\approx 4$ per CCD. As summarized in Table~\ref{tab:X}, a total of 17
confirmed group members were detected, including the largest
elliptical, NGC\,2563 itself. This galaxy is coincident with the group
center as defined by the peak of the diffuse group X-ray emission. As
it is not straightforward to cleanly separate the emission of this
galaxy from that of the intragroup medium, the central galaxy is
treated as a special case for the remainder of the paper.

\begin{deluxetable*}{llrrrrrrcccll}
\tabletypesize{\scriptsize}
\tablecaption{Group Members Detected With Chandra\label{tab:X}}
\tablewidth{0pt}
\tablehead{ \colhead{Galaxy} & \colhead{Morph.} & \colhead{$R$} &
  \colhead{log\,$L_K$} & \colhead{log\,$L_g$} &
  \colhead{Cts.} & \colhead{$L_{\rm X,th}$} &
  \colhead{$L_{\rm X,pl}$} & \colhead{$kT$} & \colhead{$\Gamma$} & 
  \colhead{$H\!R$} & \colhead{Spec.} & \colhead{Notes} \\
 \colhead{ } & \colhead{ } & \colhead{(kpc)} & \colhead{($L_\odot$)} & 
 \colhead{($L_\odot$)} & \colhead{ } & \colhead{ } & \colhead{ } & 
 \colhead{(keV)} & \colhead{ }& \colhead{ }& \colhead{ }& \colhead{ }\\
\colhead{(1)} & \colhead{(2)} & \colhead{(3)} & \colhead{(4)} & \colhead{(5)} &
\colhead{(6)} & \colhead{(7)} & \colhead{(8)} & \colhead{(9)} & \colhead{(10)}&
\colhead{(11)} & \colhead{(12)} & \colhead{(13)}  
} 
 \startdata
NGC\,2563  & E    &   0 & 11.43 & 10.54 & 1547 &  $74.2^{+5.0}_{-4.8}$ & 
  $22.6^{+5.6}_{-5.7}$ & $0.97^{+0.03}_{-0.04}$ & $2.16^{+0.21}_{-0.24}$ & 
  $0.11\!\pm\!0.01$ & 1, TP & E\\
NGC\,2562  & S0/a &  91  & 11.21 & 10.35 &   50 & $<1.87$ & 
  $2.51^{+0.58}_{-0.45}$ & 0.7$^\ast$ & $1.84^{+0.35}_{-0.30}$ & $0.56\!\pm\!0.17$ &
  2, P & E\\ 
NGC\,2560  & S0/a &  218 & 11.11 & 10.21 &  139 & $<4.49$  & 
 $8.45^{+1.65}_{-1.56}$ & 0.7$^\ast$ & $1.38^{+0.29}_{-0.27}$ & $0.59\!\pm\!0.10$ & 
  2, P & E\\
CGCG119-069 & E   &  238 & 10.13 &  9.49 &    9 & $<5.06$  & 
  $<1.39$ & 0.7$^\ast$ & 1.7$^\ast$ & $0.63\!\pm\!0.42$ & 3, U & P?\\
UGC\,04344 & Sc   & 238 & 10.14 &  9.91 &   17 & $<2.07$  & 
  $<1.86$ & 0.7$^\ast$ & 1.7$^\ast$ & $0.38\!\pm\!0.21$ & 3, U & E?, SF\\
UGC\,04332 & Sapec&  264 & 10.94 & 10.03 &  117 & $<3.82$  & 
  $6.45^{+1.18}_{-1.06}$ & 0.7$^\ast$ & $1.85^{+0.27}_{-0.27}$  & $1.32\!\pm\!0.25$
  & 2, P & E, Sy2\\
NGC\,2569  & E    &  307 & 10.71 &  9.99 &   18 & $<1.99$  & 
  $<2.09$ &  0.7$^\ast$ & 1.7$^\ast$  & $0.26\!\pm\!0.15$ & 3, UT & E?\\  
UGC\,04329 & Scpec&  439 & 10.72 & 10.06 &   57 & $<4.03$   & 
  $5.93^{+1.92}_{-1.46}$ &  0.7$^\ast$ & $2.39^{+1.16}_{-0.68}$ & $0.62\!\pm\!0.17$
  & 2, P & E, SF\\ 
IC\,2293  & SBbc &   470 & 10.37 &  9.89 &    9 & $<1.29$  & 
  $<1.00$  & 0.7$^\ast$ & 1.7$^\ast$  & $0.20\!\pm\!0.18$ &3, UT & P?, SF\\
2MJ082236 & S0/a &  542 & 9.99 & 9.32 & 13 & $<1.57$  & 
  $<1.53$  &  0.7$^\ast$ & 1.7$^\ast$  & $0.49\!\pm\!0.28$ & 3, U & E?, SF\\
NGC\,2557 & SB0 &  567 & 11.15& 10.29 &     162 & $11.2^{+3.0}_{-2.3}$ & 
  $10.4^{+1.4}_{-1.2}$ & $0.34^{+0.10}_{-0.07}$ & 1.7$^\ast$ & $0.27\!\pm\!0.05$ & 
  1, TP & E, LI\\
UGC\,04324 &  Sab &  664 & 10.71 &  9.92 &   35 & $<3.63$  & 
  $<3.96$ &  0.7$^\ast$ & 1.7$^\ast$ & $0.42\!\pm\!0.16$ & 3, U & E, SF\\ 
NGC\,2558  & Sb   &  739 & 11.03 & 10.26 &   45 &  $<2.45$  & 
  $3.90^{+0.86}_{-0.75}$ & 0.7$^\ast$ & $1.53^{+0.30}_{-0.33}$   & $0.45\!\pm\!0.15$
  & 2, P & E, LI\\ 
IC\,2338   & SBcd &  848 & 10.27 &  9.87 &   59 & $<3.09$  & 
  $4.75^{+1.02}_{-0.95}$ & 0.7$^\ast$ & 1.7$^\ast$  & $0.35\!\pm\!0.10$ & 
  2, TP& E, SF\\ 
IC\,2339&SBcpec& 858& 10.48& 10.02&          32 & $1.85^{+0.91}_{-0.84}$ & 
  $1.69^{+0.67}_{-0.60}$ & 0.7$^\ast$ & 1.7$^\ast$ & $0.21\!\pm\!0.10$ & 
  1, TP& E, SF\\
CGCG119-047 & Sab & 923 & 10.62 & 10.04 &   26 & $<2.80$  & 
 $<2.68$ &   0.7$^\ast$ & 1.7$^\ast$  & $0.36\!\pm\!0.16$ & 3, U & P?, SF\\
IC\,2341 & E/S0 &  929 & 10.91 & 10.19   &   32 & $<3.54$  & 
 $<2.89$ & 0.7$^\ast$ & 1.7$^\ast$  & $0.88\!\pm\!0.31$ & 3, U & E, LI
\enddata
\tablecomments{Col.~(3): Projected distance from peak of the
  intragroup X-ray emission. Col.~(6): Net counts in the ''full''
  0.3--7~keV band. Col.~(7): 0.5--2~keV thermal luminosities
  ($10^{39}$~erg~s$^{-1}$). Col.~(8): 0.5--2~keV power-law
  luminosities ($10^{39}$~erg~s$^{-1}$). Col.~(9): Best-fit X-ray
  temperature; asterisk indicates a fixed value. Col.~(10): Best-fit
  power-law index; asterisk indicates a fixed value. Col.~(11):
  (2--7~keV)/(0.3--2~keV) hardness ratio.  Col.~(12): Classification
  of the X-ray spectrum according to statistical quality (1--3, as described
  in Section~\ref{sec,spec}), and according to consistency with a power-law only
  (P), both thermal and power-law components present (TP), or spectral
  composition unknown (U) but hardness ratio suggests thermal
  component present (T). Col.~(13): Specification of
  whether the X-ray emission is extended (E), point-like (P), or
  uncertain (``?''), and whether the galaxy is star-forming (SF), a
  LINER (LI), or Seyfert~II (Sy2).}
\end{deluxetable*}

For all detected group members, source spectra and associated response
files were extracted within circular regions extending to where the
galaxy surface brightness becomes consistent with the local background
level. Background spectra were extracted in surrounding annuli, and
results were fitted in the 0.3--7.0~keV band using {\sc xspec}
v.\,12.3 \citep{arna96}. As many of the sources have few counts, the
maximum likelihood--based C-statistic \citep{cash79} was used to
determine best-fit parameters. The goodness-of-fit was verified using
the $\chi^2$ statistic for all sources with sufficient photon
statistics to allow meaningful constraints from spectra accummulated
in bins of at least 20~net~counts.

The X-ray emission in luminous early-types is generally dominated by
that of hot gas and low-mass X-ray binaries (e.g., \citealt{fabb06}).
We modeled any hot gas emission in the group members using the {\em
  mekal} optically-thin thermal plasma model \citep{mewe85,lied95} in
{\sc xspec}, assuming solar abundance ratios. Any X-ray binary/active
galactic nucleus (AGN) component was modeled with a power law. In all
fits, the absorbing hydrogen column density was fixed at the Galactic
value ($N_{\rm H}=4.0\times10^{20}$~cm$^{-2}$; \citealt{kalb05}),
since the limited photon statistics generally precluded useful
constraints on $N_{\rm H}$ from spectral fitting itself. Following
\citet{sun07} and \citet{jelt08}, the metallicity in the {\em mekal}
model was fixed at $0.8~Z_\odot$, the mean value found for galactic
hot gas in the large cluster galaxy sample of \citet{sun07}. This left
four free model parameters: Gas temperature $T$, power-law index
$\Gamma$, and two normalizations. However, due to limited number of
counts, the $3\sigma$ uncertainties on $T$ and $\Gamma$ were
unconstrained for most of the galaxies. In such cases we first fixed
$\Gamma$ at 1.7, as is typical for X-ray binary and AGN spectra, and,
if necessary, also $T$ at 0.7~keV. Based on the photon statistics and
spectral results, the detected galaxies were then divided into three
categories, as summarized in Col.~(12) of Table~\ref{tab:X}:

(1) Galaxies for which the normalization of both components was
well-constrained at the $2\sigma$ level (three galaxies in total;
well-constrained here means that the $2\sigma$ uncertainties are
finite, and that the normalization is non-zero within those
uncertainties). For the central galaxy NGC\,2563, all of the four
parameters are well constrained, but its emission is not easily
deblended from that of the intragroup medium. The other two galaxies
are NGC\,2557, where $\Gamma$ was fixed, and IC\,2339, where also $T$
was fixed. Best-fit values and $1\sigma$ errors for all fit parameters
were determined using Markov Chain Monte Carlo (MCMC) simulations in
{\sc xspec}. The values reported were obtained from the median value
of the chain. We note for completeness that the standard deviation
leading to a parameter being here considered well-constrained (which
is based on the change in the adopted fit statistic when the parameter
is varied, using {\sc xspec}'s standard ``error'' command) is not
necessarily identical to that resulting from the MCMC simulations.

(2) Galaxies with $\ga 30$~counts but for which the normalization of
one or both components remained unconstrained within the $2\sigma$
uncertainties (six galaxies in total). In the cases where the
$2\sigma$ lower limit on the thermal luminosity $L_{\rm X,th}$ was
consistent with zero, whereas that on the power-law component was not,
the spectrum was assumed to be consistent with a power-law
only. IC\,2338 is consistent with displaying both a thermal and a
power-law component, but, unlike galaxies in category~(1), its
$2\sigma$ upper limit on $L_{\rm X,th}$ remained unconstrained. Using
MCMC, we report the corresponding $1\sigma$ upper limit.  For the
remaining galaxies in this category, the best-fit $\Gamma$ was
consistent with $\approx 1.7$ as expected from X-ray binaries or an
AGN, whereas the best-fit temperature was much higher ($T >4$~keV)
than expected for hot galactic gas. Hence, a power-law component
likely dominates their X-ray output. For these sources, we fixed the
best-fit power-law model and added a $T=0.7$~keV thermal
component. The limit on $L_{\rm X,th}$ was determined from the
$1\sigma$ upper limit of the normalization of the {\em mekal} model.

(3) Galaxies detected with $\la 30$~net counts (eight in total).  In
these cases, statistics are insufficient to allow a robust
determination of the nature of their X-ray emission.  A $1\sigma$
upper limit to their $L_{\rm X,th}$ was estimated assuming a
$T=0.7$~keV {\em mekal} model, but these galaxies are labeled with
``U" (for unknown spectral composition) in Col.~(12) of
Table~\ref{tab:X}. To help clarify the nature of these sources, we
also considered their hardness ratios $H\!R$ (here the
2--7~keV/0.3--2~keV flux ratio). This ratio is 0.02--0.05 for a {\em
  mekal} model with $T=0.7$--1.0~keV, and 0.35--0.47 for a power-law
spectrum with $\Gamma=1.7$--2.0. For $0.05\!<\!H\!R\!<\!0.35$, the
dominant component is ambiguous, but some contribution from a thermal
component is allowed. Although these classifications are only
tentative, given the Poisson uncertainties on $H\!R$, we do note that
all group members spectroscopically confirmed to require only a
power-law show $H\!R>0.4$, whereas $H\!R\lesssim0.35$ for the galaxies
with a spectroscopically identified thermal component. In
Table~\ref{tab:X}, a ``T" is listed after ``U" in Col.~(12) if
$H\!R<0.35$, to indicate that the galaxy may contain thermal emission.
The remaining galaxies are labeled with either ``TP", meaning that
both thermal and power-law components are likely present
($0.05\!<\!H\!R\!<\!0.35$), or ``P", indicating consistency with a
power-law only ($H\!R>0.35$).

To further aid in identifying the nature of galactic X-ray emission,
we used its spatial extent to help discriminate between nuclear (AGN)
and galaxy-wide emission (e.g., hot gas halos, X-ray binaries).  For
each detected group member, we fitted the 0.5--2~keV brightness
profile with a Gaussian, and classified it as extended if the
$1\sigma$ lower bound of its full width at half-maximum (FWHM) was
$\geq 10$\% larger than that of the local point spread function (PSF)
extracted at peak source energy. Overall, most of the detected group
members appear extended in X-rays, and these are labeled with an ``E''
in Col.~(13) of Table~\ref{tab:X}, or with a ``P'' (point-like)
otherwise. For most sources with $\la 30$~counts the extent remains
poorly constrained, however, and Col.~(13) then includes a question
mark.

For the X-ray undetected group members, we followed the procedure of
\citet{sun07} and \citet{jelt08} to estimate luminosity upper
limits. Count rates were measured within a circular aperture of $R=
3$~kpc from the optical galaxy center, in a few cases expanded to
include the $90\%$ encircled energy radius of the local PSF at
$E=1$~keV to account for PSF smearing. The upper limit on $L_{\rm
  X,th}$ was derived from the Poisson $3\sigma$ upper limit on the
count rate \citep{gehr86} assuming a $T=0.7$~keV {\em mekal} model.

\vspace{2mm}

\subsection{VLA Observations and Analysis}\label{sec,hiobs}

The H{\sc i} analysis was based on two different observing runs at the
VLA in its most compact, 1~km (D-array), configuration. The first run,
done in 1999, covered the entire group with a $6\times 6$ point
mosaic. Individual pointings were separated by $15'$, fully sampling
the primary beam with its FWHM of $30\arcmin$. To probe as wide a
velocity range as possible with sufficient velocity resolution, we
used a total bandwidth of 3.125~MHz and four Intermediate Frequency
(IF) channels with two frequency settings and slightly overlapping
bands. The total velocity range covered was $\sim$\,1100~km~s$^{-1}$
with a resolution of 20~km~s$^{-1}$. The data were calibrated using
standard AIPS\footnote{http://www.aips.nrao.edu/} procedures, and then
imported into
MIRIAD\footnote{http://www.atnf.csiro.au/computing/software/miriad/}
for mosaicing and joint deconvolution, turning the individual
pointings into one mosaiced cube for each IF. The continuum was then
subtracted from the $u$--$v$ data using a fit through the line-free
channels identified within the cube. However, inspection of the final
cube showed that our velocity coverage was insufficient, with several
group members detected close to the edge of the band and their H{\sc
  i} velocity range only partly probed.

To remedy this, NGC\,2563 was reobserved in 2007, using a 14-point
mosaic that covered an area identical to that of our {\em Chandra}
mosaic. To also cover a larger velocity range, we used a total
bandwidth of 6.25~MHz with two IFs and no online Hanning
smoothing. Each pointing was observed at two separate frequencies,
resulting in a total velocity coverage of 2100~km~s$^{-1}$.  Each
pointing and velocity setting was calibrated separately, and cubes
were made using AIPS. Overlapping frequencies were then combined in
the image plane, resulting in a 101--channel cube per pointing. After
identifying channels containing H{\sc i}, the continuum was
iteratively subtracted from each cube using a linear fit in the image
plane through the line--free channels.  The resulting cubes were then
CLEANed, combined into one, and the output corrected for the primary
beam response. Where cubes overlapped, they were averaged with
weighting that accounted for the location within the pointing and for
the distance from the center of each field. Although interference
generated by the VLA rendered data unusable within a narrow velocity
range (4331--4374~km~s$^{-1}$), the only group member affected is an
Sm dwarf with $L_K \approx 1\times 10^9 L_\odot$, so this has little
impact on our results.

Instrumental parameters and specifics of the cubes for both data sets
are summarized in Table~\ref{tab,VLA}, including their velocity ranges
(heliocentric, optical definition). Since complete velocity coverage
is important here, the 36-point mosaic was used exclusively to search
for H{\sc i} emission, combining it with the central part of the
14-point mosaic. All remaining H{\sc i} analysis employed the 14-point
mosaic only. This was searched for neutral hydrogen using three
different methods. First the 101--channel cube was divided into groups
of 25, smoothed spatially to half resolution, and Hanning--smoothed in
velocity. This was then used as a mask for the full-resolution cube,
blanking all pixels below $2\sigma$ in the smoothed cube and summing
over 25~channels. Next we searched the cube by eye, stepping through
the velocity channels at different speeds and comparing our detections
with the first method. Finally, we plotted spectra integrated over a
few beams at the location of the optically identified group
members. This did not give any additional detections, so our search
was essentially an optically blind one.

\begin{table}
\begin{center}
\caption{Parameters of the VLA D-array Observations\label{tab,VLA}}
\begin{tabular}{lcc} \tableline \hline
Parameter  &  36-Point Mosaic  & 14-Point Mosaic \\ \hline
RA (J2000) & 08 20 23.7 &  08 20 23.7  \\
Dec (J2000) & 21 05 00.5 & 21 05 00.5 \\
Vel.~range (km~s$^{-1}$) & 4364--5449 & 3834--5942 \\
Vel.~resolution (km~s$^{-1}$) & 21   & 21 \\
Synthesized beam ($\arcsec$) & $70\times 59$ & $64\times 53$ \\
{\em rms} noise (mJy beam$^{-1}$)  &   0.5 & 0.6 \\ \hline
\end{tabular}
\end{center}
\end{table}

\begin{figure*}
\begin{center}
\epsscale{.8}
\plotone{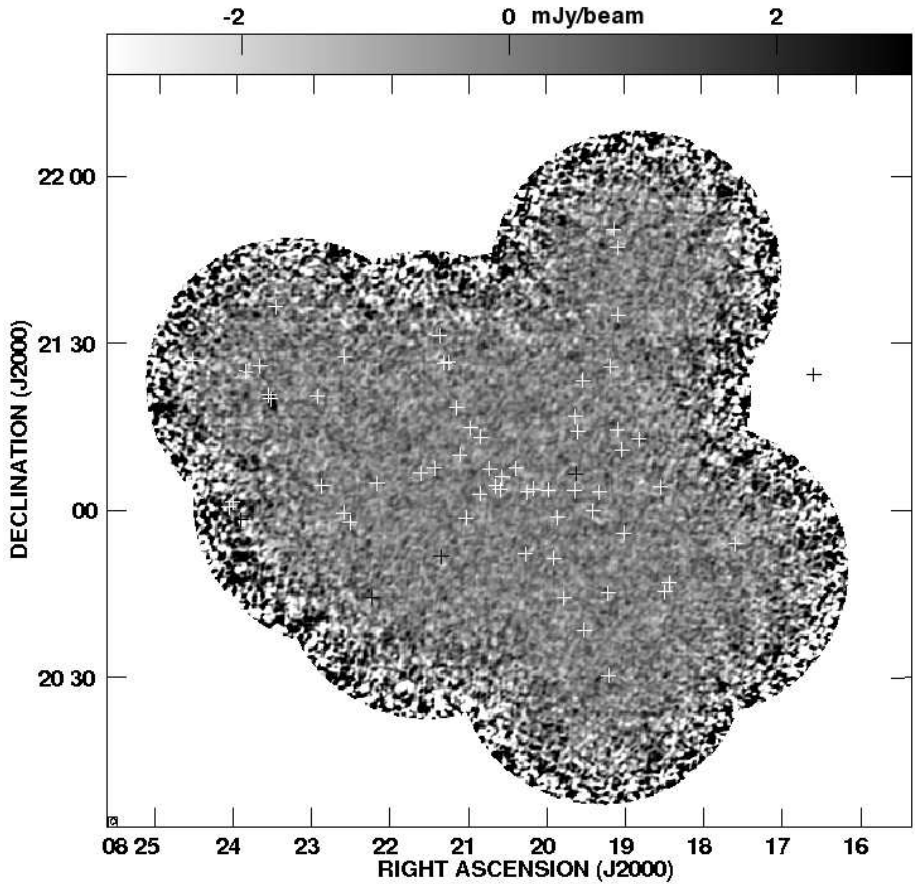}
\end{center}
\figcaption{Grey-scale image of a typical channel in the H{\sc i} data
  cube for the 14 point VLA mosaic. Crosses mark the position of the
  group members. The one group member not covered in the 14 point VLA
  mosaic, IC\,2253, was observed but not detected in the 36 point
  mosaic. 
\label{fig:HIchannel}}
\end{figure*}

Figure~\ref{fig:HIchannel} shows a grey-scale image of a typical
channel in the cube, to illustrate the spatial noise distribution and
the position of the group members. In the central square degree ($R
\la 600$~kpc~$\approx 0.7 R_{\rm vir}$) the {\em rms} noise (using
robust weighting) is remarkably uniform at 0.6~mJy~beam$^{-1}$,
equivalent to $N_{\rm H} \simeq 4 \times 10^{18}$~cm$^{-2}$. Assuming
$3\sigma$ over five~channels, the corresponding sensitivity is $M_{\rm
  HI} = 2 \times 10^8$~M$_{\odot}$. Along the $9\arcmin$ wide outer
edges, the noise and sensitivity rise to 3.0~mJy~beam$^{-1}$ and
$1\times 10^9$~M$_{\odot}$, respectively, due to the correction for
the primary beam response. This only affects five out of the 64 group
members and none of our conclusions. A total noise--free H{\sc i}
image was produced by again using the smoothed cube as a mask and
summing each detected galaxy over the narrow velocity range in which
H{\sc i} was detected. The total H{\sc i} mass for each galaxy was
then determined by summing the flux in individual channels in a box
centered on the galaxy. The resulting typical uncertainty is 10\% for
the galaxies with $M_{\rm HI}$ above a few times $10^9$~M$_{\odot}$,
and $2\times 10^8$~M$_{\odot}$ for the remainder.

\vspace{3mm}

\section{Results}\label{sec,results}

\subsection{X-ray Results}\label{sec,X}

Table~\ref{tab:X} lists all 17 X-ray detected group members (of the 54
observed), and whether these show evidence for star formation or AGN
activity. This is based on their location in a BPT diagram
\citep{bald81,kewl01,kauf03b}, as inferred from an emission-line
analysis of their SDSS spectra, with optical line fluxes measured
using the code described in \citet{trem04}.

\begin{figure*}
\begin{center}
\epsscale{.95} 
\plotone{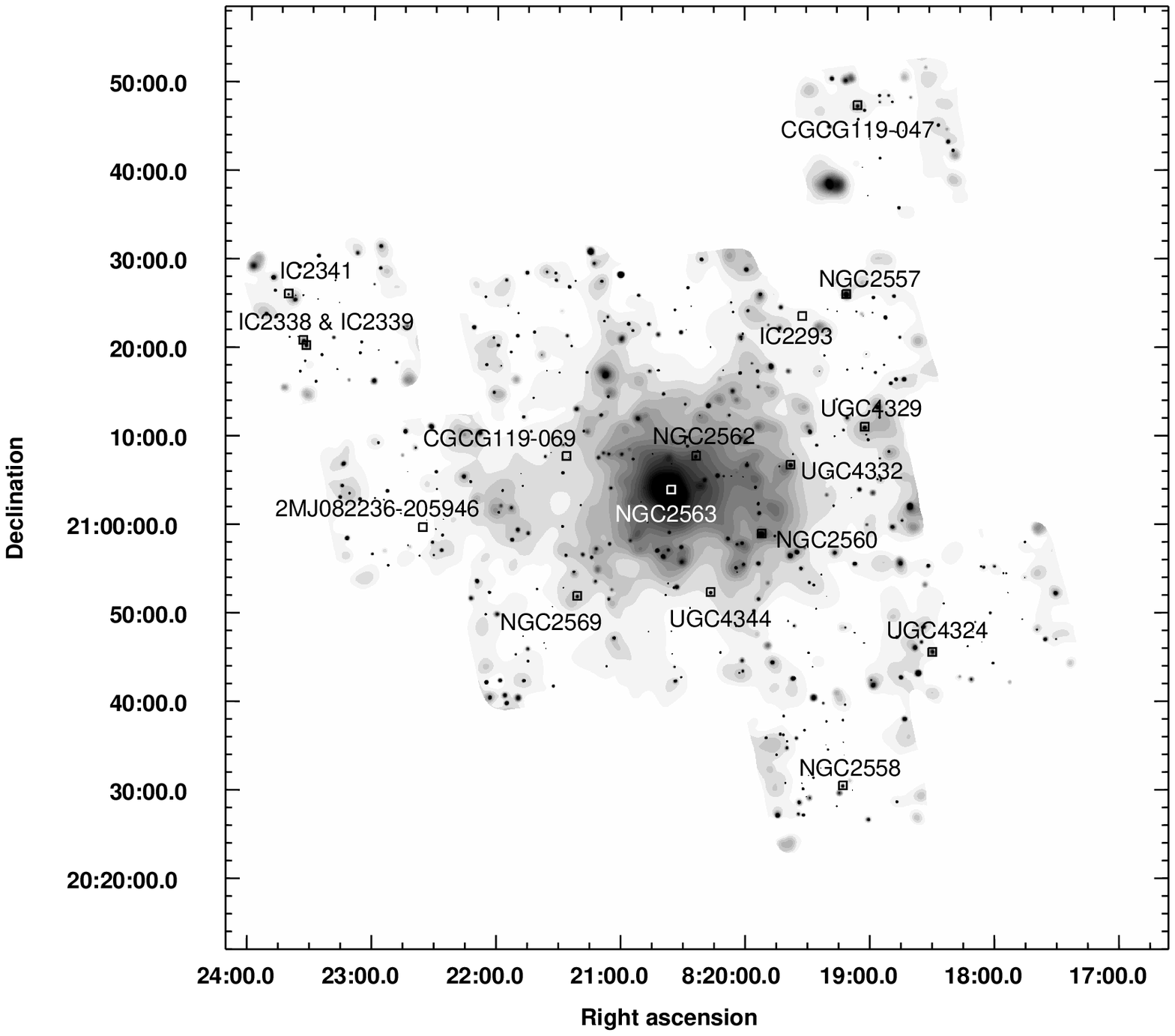}
\end{center}
\figcaption{Adaptively smoothed 0.5--2.0~keV image of the NGC\,2563
  group from our 14 {\em Chandra} ACIS-I pointings, with intensity
  plotted on a log scale. All group members detected in the unsmoothed
  {\em Chandra} data are marked with squares.
  \label{fig:mosaic}}
\end{figure*}

Figure \ref{fig:mosaic} shows a 0.5--2~keV mosaic image of the group
from our 14 {\em Chandra} pointings, adaptively smoothed using the
'csmooth' algorithm, with signal significance set between
3$\sigma$--5$\sigma$. To produce this, spectrally weighted exposure
maps were smoothed to the same spatial scales, and the
exposure-corrected images combined into one. The impact of edge
effects was reduced by removing regions with (smoothed) exposure
values below 3\% of the maximum, but small enhancements at the
boundaries of adjacent pointings still remain. Detected group galaxies
are marked with squares in the figure, with interlopers accounting for
the remaining X-ray sources.

Due to the spatial variation in {\em Chandra} effective area, our
X-ray detection limits are not uniform across the field, but we reach
a $3\sigma$ upper limit in the 0.5--2.0~keV band of $L_{\rm X} \approx
3\times10^{39}$~erg~s$^{-1}$ or better for all of the observed group
members. These span almost three orders of magnitude in $L_K$, with
nearly all galaxies above $0.5 L^\ast$ in the $g$-- or $K$--band
detected in X-rays (7/8 in $K$, 15/16 in $g$), whereas the
corresponding fraction is less than half (4/9) for those fainter than
$0.2L_K^\ast$. This is qualitatively consistent with other studies
which show a strong correlation between galaxy optical and X-ray
luminosity \citep{fabb92,read01,sull01,sull03}.  As noted earlier, the
ten members not observed by {\em Chandra} are mostly optically faint,
with seven below $0.1L_{K}^\ast$ and only one above $L_{K}^\ast$.

\begin{figure}
\begin{center}
\epsscale{1.17}
\plotone{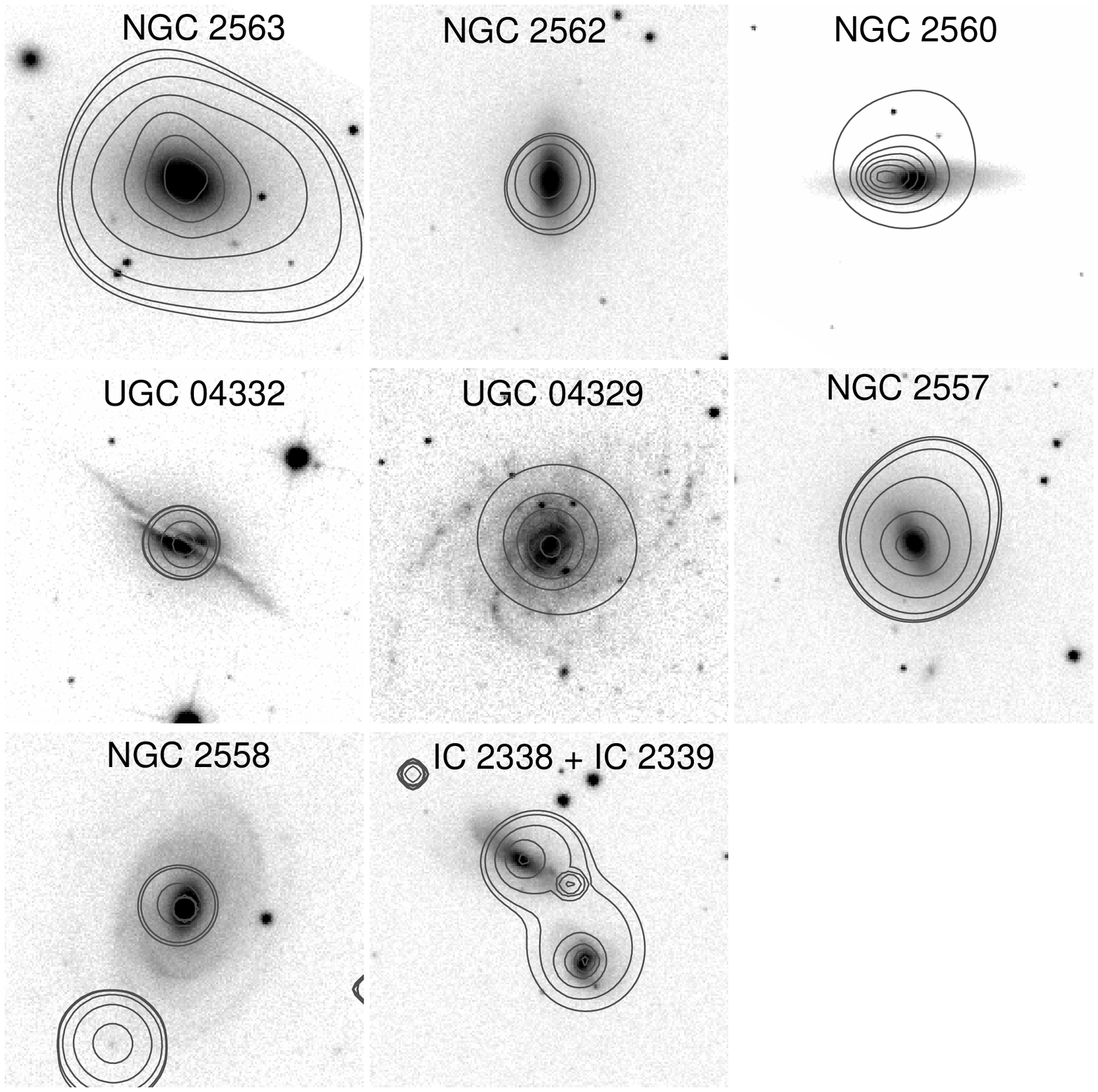}
\end{center}
\figcaption{Smoothed 0.5--2 keV X-ray contours overlayed on SDSS
  $r$-band images for the eight X-ray brightest group members. Each
  image is $2\arcmin \times 2\arcmin$ ($\sim 40 \times 40$~kpc).
  \label{fig:2}}
\end{figure}

Among the nine galaxies for which X-ray spectral fitting provides
useful constraints (cases~1 and 2 in Section~\ref{sec,spec}), four
show evidence for a thermal component, including the central NGC\,2563
itself. From the hardness ratios of the remaining (case~3) galaxies,
two additional objects show evidence for thermal emission, for a total
of six such galaxies. However, recalling from Section~\ref{sec,spec}
that $H\!R < 0.05$ for a thermal--only model, we note that no source
besides NGC\,2563 itself has a hardness ratio $H\!R<0.2$, so a
power-law component is likely present (if not dominant) in all the
X-ray detected galaxies. In all group members with more than
40~counts, the emission is inferred to be spatially extended, and
their smoothed 0.5--2~keV contours overlayed on SDSS $r$-band images
are shown in Figure~\ref{fig:2}. Note the offset between the X-ray
peak and the optical center in NGC\,2560, possibly indicating the
presence of an ultra-luminous X-ray source as seen in other nearby
galaxies \citep{mill04,robe07}.

Previous studies have revealed a correlation between $L_K$ and thermal
X-ray luminosity for early-type galaxies in all environments
\citep{sull01,sun07,jelt08,mulc10}. Taking early-types to be Sa or
earlier for consistency with these works, then roughly half of the
members in NGC\,2563 belong to this category.
Figure~\ref{fig,scaling}(a) compares $L_{\rm X,th}$ for our optically
bright group members to the scaling relation derived by \citet{jelt08}
for early-types in groups. The central galaxy NGC\,2563 itself is here
excluded for reasons discussed earlier. Our two other early-types with
evidence for a thermal component (NGC\,2557 and NGC\,2569) are
consistent with the \citet{jelt08} relation, while the majority of the
X-ray undetected galaxies are too optically faint to offer much
further insight with the present X-ray detection limits.  With the
limited statistics, we simply conclude that the detected early-types
in NGC\,2563 do not deviate significantly from the $L_{\rm X}$--$L_K$
relation found for other group ellipticals.

\begin{figure}
\begin{center}
\epsscale{1.17}
\plotone{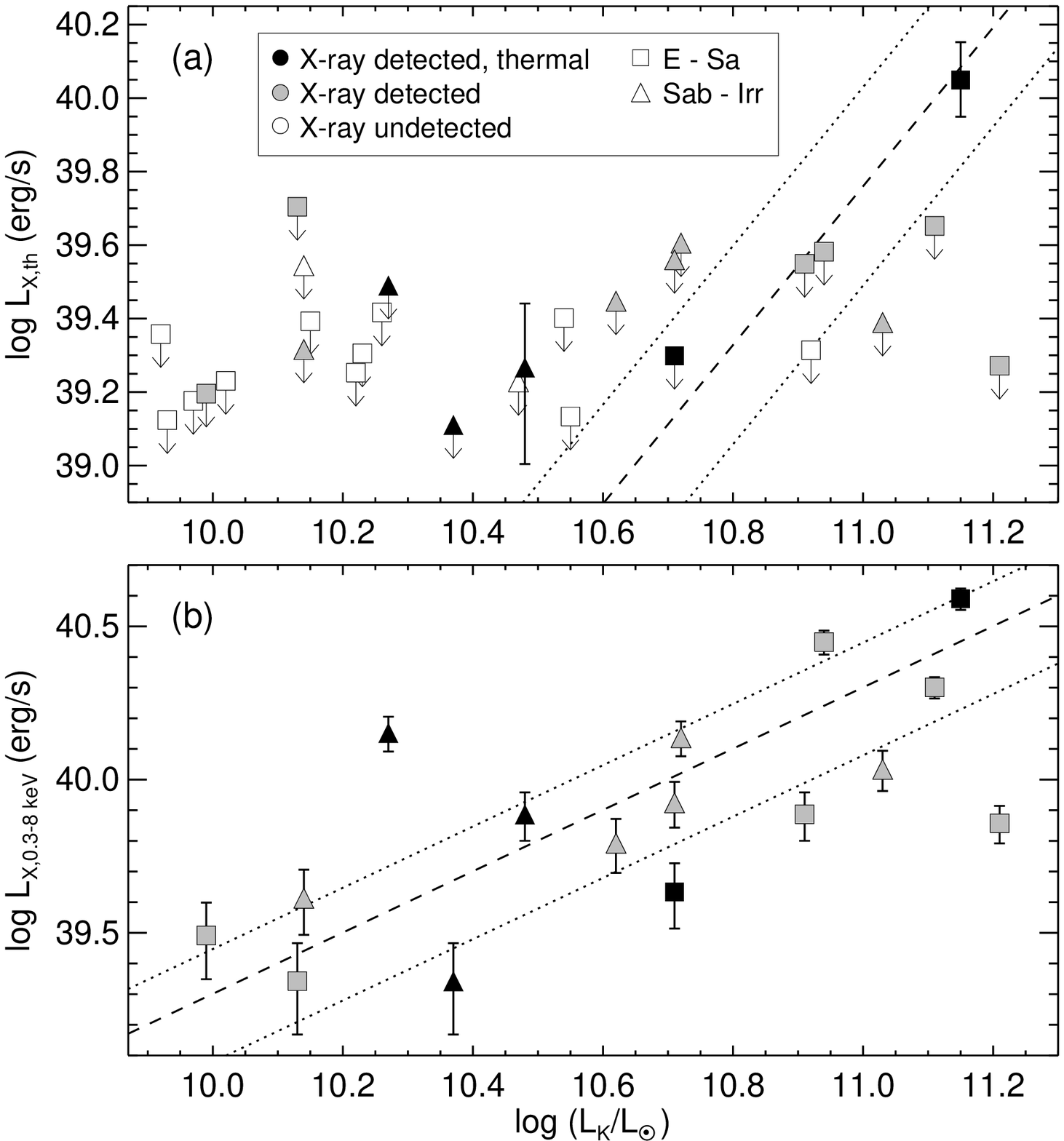}
\end{center}
\figcaption{(a) Thermal X-ray luminosities of all optically bright
  group members covered by {\em Chandra} (excluding the central
  NGC\,2563 itself), and the corresponding scaling with galaxy $L_K$
  for early-types in groups from \citet{jelt08} (dashed) and its
  $3\sigma$ errors (dotted). Black symbols represent galaxies with
  evidence for a thermal component, grey symbols the remaining X-ray
  detected galaxies, and empty symbols the X-ray undetected
  objects. Early-type galaxies are shown by squares, late-types by
  triangles. (b) 0.3--8~keV $L_{\rm X}$ for all X-ray detected
  members, again excluding NGC\,2563 itself, along with the scaling
  relation (and $1\sigma$ errors) for total $L_{\rm X}$ from low-mass
  X-ray binaries in early-types from \citet{kim04}. Symbols are as
  above.
  \label{fig,scaling}}
\end{figure}

To further elucidate the potential hot gas contribution to the total
$L_X$ of our early-types, Figure~\ref{fig,scaling}(b) compares the
0.3--8~keV $L_X$ of the X-ray detected group members to that expected
from low-mass X-ray binaries using the scaling relation of
\citet{kim04}. NGC\,2563 itself is again excluded. None of of our
early-types shows evidence of a significant X-ray excess relative to
this expectation, suggesting that the majority of their X-ray output
is of stellar origin. A few of the more luminous galaxies even lie
below this relation, including NGC\,2557 which does contain a thermal
component.

Of the eight X-ray detected late-type members, three show evidence for
a thermal component. All eight galaxies have $L_g>0.5L_{g}^\ast$ but
span a relatively larger range in $L_{K}$, suggesting their total
X-ray output is more directly related to the blue stellar light than
to stellar mass. We note the similarity to other studies of late-type
galaxies that find their diffuse $L_{\rm X}$ to correlate more
strongly with $L_B$ (and hence star formation rate) than with $L_K$ or
stellar mass \citep{tull06,sun07}. Our SDSS spectra generally support
this, with seven of our eight X-ray detected late-type members showing
optical line emission consistent with star formation. The eighth
galaxy (NGC\,2558) has line ratios consistent with a LINER. Based on
the BPT diagram, there are 12 other actively star-forming late-types
in the group, but all have $L_g < 0.5L_g^{\ast}$ and remain X-ray
undetected.

\subsection{HI Results}\label{sec,HI}

We detect H{\sc i} in 20 of the 64 group members down to our H{\sc i}
mass limit of $\sim$ $2\times 10^8$~M$_{\odot}$ (recall that our
limits are higher for galaxies at the very edge of the VLA
mosaic). The results for these 20 objects are summarized in
Table~\ref{tab:HI}, along with results for three large spirals that
remain undetected in H{\sc i} but whose H{\sc i} deficiencies can
still be estimated (see below). Quoted H{\sc i} velocity widths
correspond to the velocity range of channels with signal at the $\geq
2\sigma$ level; note that for UGC\,04332 the H{\sc i} profile extends
to the edge of our velocity coverage, so the H{\sc i} width and mass
provided in Table~\ref{tab:HI} are lower limits. Two of the 20 H{\sc
  i} detections are associated with galaxies not previously identified
as group members from optical spectroscopy (SDSSJ082044.60+210715.0
and SDSSJ081931.17+203843.9). Both of these represent optically faint
irregular galaxies.

\begin{table*}
\begin{center}
\caption{Parameters of the VLA D-array Observations\label{tab:HI}}
\begin{tabular}{llrrccrr} \tableline \hline
Galaxy & Morph. & $R$ & H{\sc i} width & H{\sc i} vel. &
  Optical vel. & Log\,(M$_{\rm HI}$) & Def$_{\rm HI}$ \\
  &  & (kpc) & (km~s$^{-1}$) & (km~s$^{-1}$) & (km~s$^{-1}$) & (M$_{\odot}$) \\ 
\hline
SDSSJ082044.60+210715.0& Irr& 73 & 42 & 4406\tablenotemark{a} & -- & 8.28 & -- \\ 
CGCG119-061& Sab & 112 & -- & -- & 5235 & $<$8.60\tablenotemark{b} & $>$$+0.56$ \\
CGCG119-059&Sc & 165 & 170 & 4215 &4211 & 8.73 & $+0.39$ \\
UGC\,04344&Sc    & 238 & 149 & 5033 & 5041 & 9.89 & $-0.10$ \\
SDSSJ081938.81+210353.0 &Irr & 254 & 43 & 4236 & 4240 & 8.48 & -- \\
UGC\,04332 & Sa pec & 264 & $>$449\tablenotemark{c}  & 5481 & 5514 & 
$>$9.41\tablenotemark{c}   & $<$$+0.04$\tablenotemark{c}  \\
CGCG119-053 & Sa pec & 339 & 171 & 4852 & 4877 & 9.26 & $-0.17$ \\
UGC\,04329 & Sc pec & 439 & 254 & 4109 & 4099 & 9.98 & $+0.10$ \\
SDSSJ081905.46+211448.0 & Sm? & 451 & 106 & 4842 & 4844 & 8.51 & -- \\
IC\,2293& SBbc & 470 & 212 & 4088 & 4094 & 8.91 & $+0.55$ \\
CGCG119-043& Sc pec & 506 & 170 & 4470 & 4458 & 8.65 & $+0.47$ \\
CGCG119-051& Sb\tablenotemark{d}  & 513 & 191 & 5033 & 5028 & 9.32 & $-0.02$ \\
SDSSJ081931.17+203843.9& Irr & 564 & 85 & 4980\tablenotemark{a} & -- & 8.81 & -- \\
CGCG119-040& Sa & 663 & 21 & 4841 & 4816 & 7.78 & $+1.23$ \\
UGC04324& Sab\tablenotemark{d}  & 664 & 340 & 4831 & 4814 &  9.20 & $+0.32$ \\
SDSSJ081904.25+213521.0& Irr & 724 & 106 & 4863 & 4865 & 8.95 & -- \\
NGC\,2558& Sb & 739 & 405 & 4990 & 4998 & 10.08 & $-0.23$ \\
IC\,2338 & SBcd\tablenotemark{d} & 848 & 313\tablenotemark{e}
 & 5413\tablenotemark{e} & 5400 & 9.76\tablenotemark{e} & $-0.34$\tablenotemark{e} \\
 IC\,2339 & SBc pec & 858 & 313\tablenotemark{e} & 5413\tablenotemark{e} &5420 & 9.76\tablenotemark{e} & $-0.34$\tablenotemark{e} \\
CGCG119-082&SBa & 898 & -- & -- & 4783 & $<$9.00\tablenotemark{b}  & $>$$+0.09$ \\
UGC\,04386& Sab pec & 921 & -- & -- & 4640& $<$9.00\tablenotemark{b}  & $>$$+0.61$ \\
CGCG119-047&Sab\tablenotemark{d} & 923 & 276 & 4502 & 4506 & 9.73 & $-0.45$\\
SDSSJ082352.25+212507.2 &Sc & 964 & 171 & 5086 & 5095 & 8.85 &$+0.45$ \\ \hline
\end{tabular}
\end{center}
\tablecomments{(a) New H{\sc i} group member. (b) H{\sc i}
  undetected. (c) Incomplete H{\sc i} velocity coverage. (d) H{\sc i}
  tail. (e) Closely interacting pair. Values apply for the entire
  system.}
\end{table*}

\begin{figure*}
\begin{center}
\epsscale{.92}
\plotone{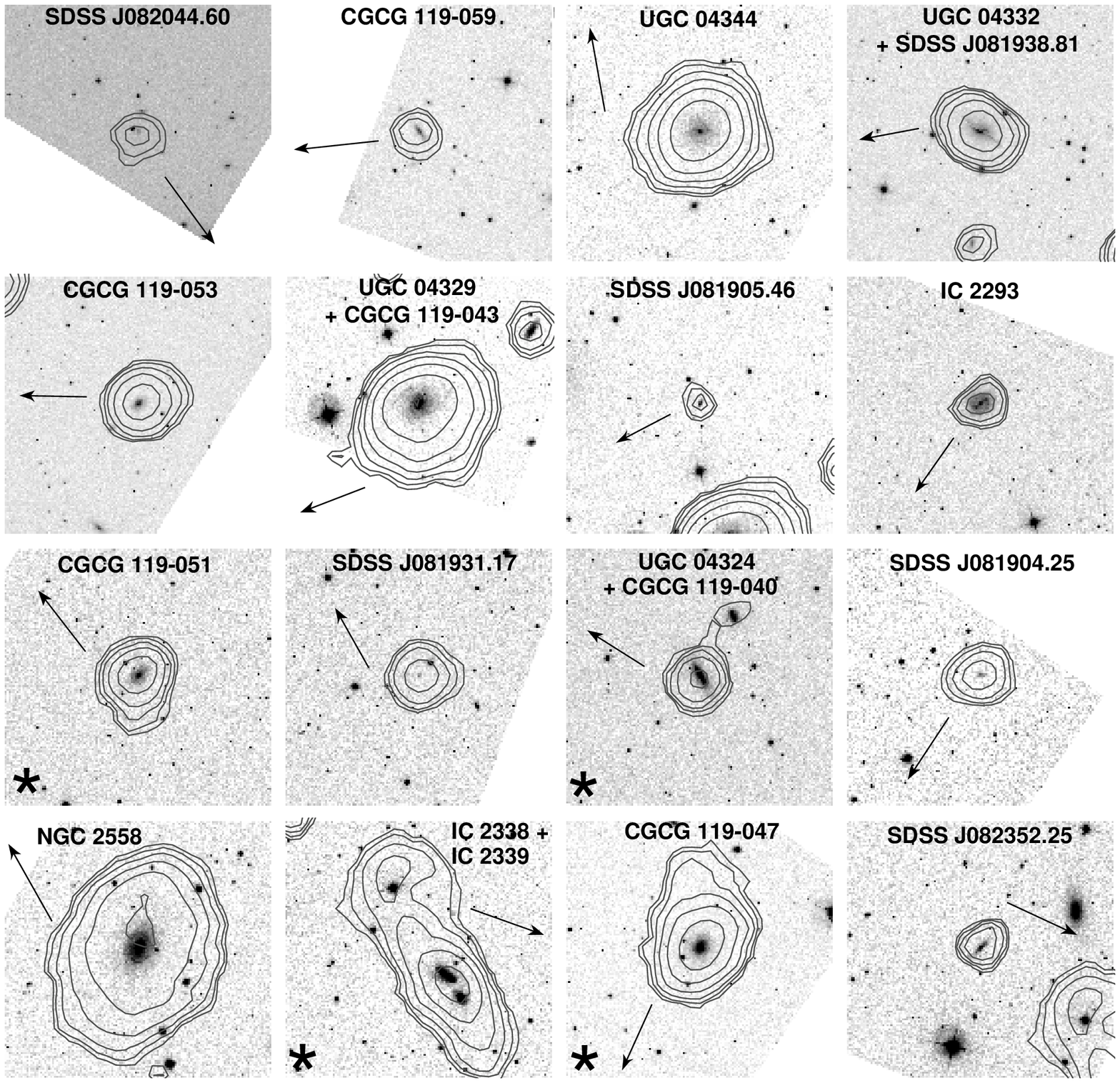}
\end{center}
\figcaption{H{\sc i} contours overlayed on SDSS $r$-band images for
  the 20 H{\sc i} detected group members, in order of increasing
  distance from the group center. Each image is $7' \times 7'$ ($\sim
  130\times 130$~kpc), with contours at 0.04, 0.08, 0.16, 0.32, 0.64,
  1.28, and 2.50~Jy~beam$^{-1}$. Arrows indicate the direction to the
  group centre. Objects with significant H{\sc i} tails/extensions are
  marked with an asterisk in the lower left corner.
\label{fig:SDSS_HI}}
\end{figure*}

H{\sc i} is detected only in galaxies of types Sa and later, including
in all 15 such galaxies with $L_K >0.1L_K^\ast$.
Figure~\ref{fig,hi_mosaic} shows contours from our noise--free H{\sc
  i} mosaic overlayed on the {\em Chandra} mosaic of the full group,
and Figure~\ref{fig:SDSS_HI} displays an H{\sc i}/optical overlay for
each H{\sc i} detected galaxy. The latter figure reveals four
significant H{\sc i} extensions in the group associated with six
galaxies (IC\,2238/IC\,2339, UGC\,04324/CGCG119-040, CGCG119-047 and
CGCG119-051, all labeled with an asterisk in the Figure). This implies
that $\sim 30$\% of our H{\sc i} detected group members display H{\sc
  i} evidence for being involved in an ongoing interaction. The
galaxies IC\,2338 and IC\,2339 represent a particularly close pair,
and the H{\sc i} quantities quoted in Table~\ref{tab:HI} therefore
apply for the entire system. The absence of detectable H{\sc i} among
the early-type group members is consistent with previous studies
finding H{\sc i} above our detection limits in only a small fraction
of early-types \citep{burs87,morg06,dise07}. Since we have performed
an optically blind search for H{\sc i} across the entire group, we can
furthermore rule out the existence of any optically dark H{\sc i}
clouds not associated with any galaxy down to our H{\sc i} mass limit.

\begin{figure*}
\begin{center}
\epsscale{.85}
\plotone{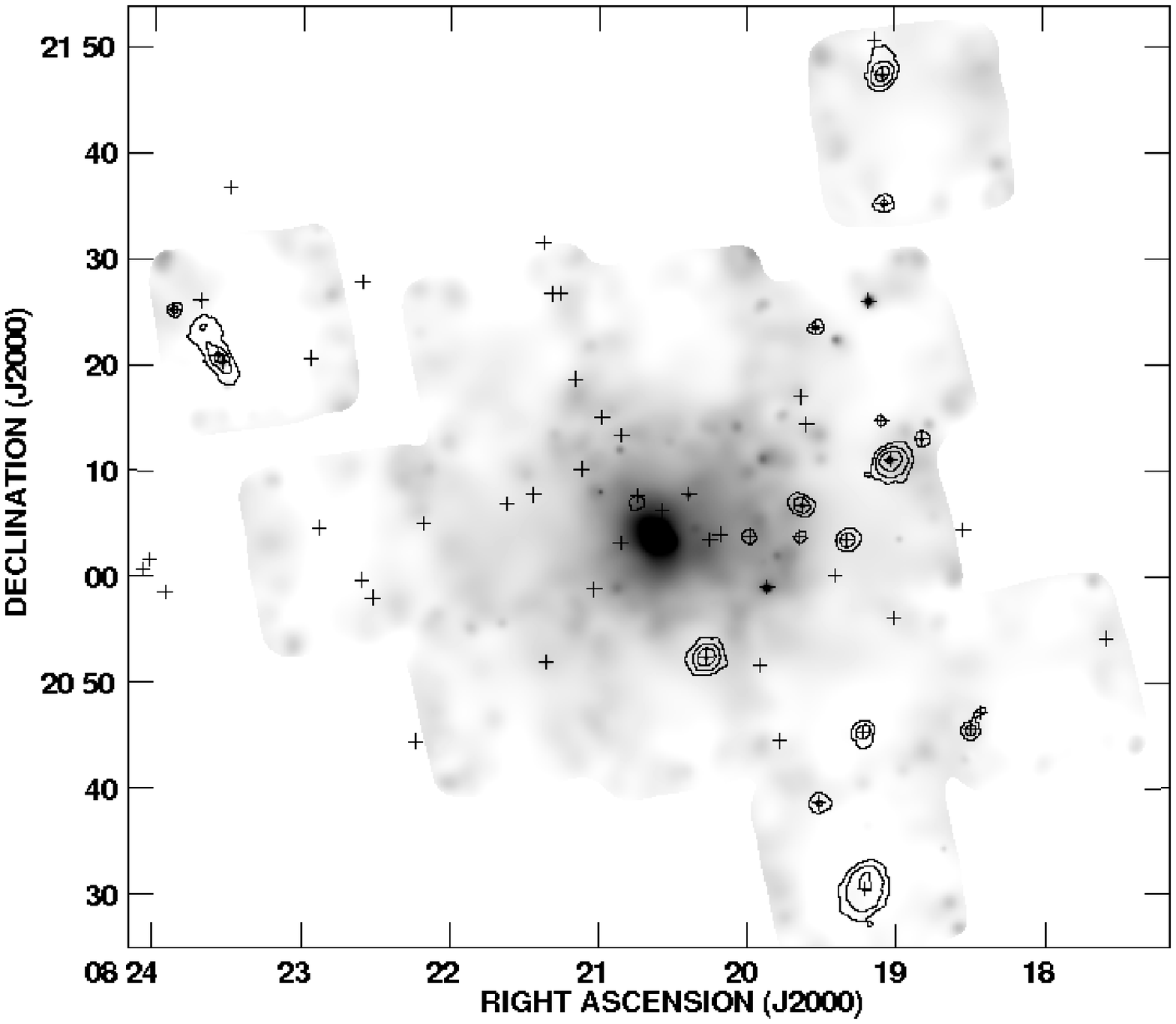}
\end{center}
\figcaption{0.5--2.0~keV grey-scale image of the group, with total
  H{\sc i} contours overlayed at 0.04, 0.4, 1.2, and 2.4
  Jy~beam$^{-1}$~km~s$^{-1}$ (corresponding to $N_{\rm H} = 1.27$,
  12.7, 38.2, and $76.4\times 10^{19}$~cm$^{-2}$). Individual group
  members are marked by crosses; IC\,2253, not covered by the 14-point
  VLA mosaic, is not included.
\label{fig,hi_mosaic}}
\end{figure*}

To investigate to what extent the group members may be deficient in
H{\sc i}, we estimated H{\sc i} deficiencies Def$_{\rm HI}$ following
the usual prescription,
\begin{equation}
\mbox{Def}_{\rm HI} = \mbox{log}\,(M_{\rm HI})_{\rm
  expected} - \mbox{log}\,(M_{\rm HI})_{\rm observed}.
\end{equation} 
The expected H{\sc i} masses were computed from the best-fit
relationship between total H{\sc i} mass and galaxy optical diameter
$D$ derived for galaxies in the field \citep{sola96}. $D$ is here
defined as the galaxy major axis measured from the Palomar Observatory
Sky Survey blue prints. Where not already available from the Uppsala
General Catalog of Galaxies \citep{nils73}, we measured $D$ directly
from these prints. As the \citet{sola96} field sample was restricted
to rather large spirals of types Sa through Sc, only group members of
those types and with $D>10$~kpc were considered here. Note, as
mentioned above, that this includes three galaxies not detected in
H{\sc i}, and these are also listed in Table~\ref{tab:HI}. For the
IC\,2338/2339 pair, a single deficiency was calculated from the total
observed and expected H{\sc i}~masses given the optical sizes of both
galaxies.

The resulting deficiencies listed in Table~\ref{tab:HI} span a wide
range, but given the uncertainties in $D$ and optical morphology, as
well as the dispersion in the H{\sc i} masses of field galaxies,
objects with observed H{\sc i} masses within a factor of a few of the
expected values cannot be considered deficient. However, there are six
galaxies with Def$_{\rm HI} \ga 0.45$, consistent with an H{\sc i}
deficiency of at least a factor of three. One of these is UGC\,04386,
but Def$_{\rm HI}$ for this object is not well-established, in part
because it resides at the eastern edge of the H{\sc i} mosaic where
the data are noisier, and in part because it is viewed edge-on and is
of uncertain morphological type.

\subsection{The Intragroup Medium}

\begin{figure}
\begin{center}
\epsscale{1.2}
\mbox{
\hspace{1mm}
\plotone{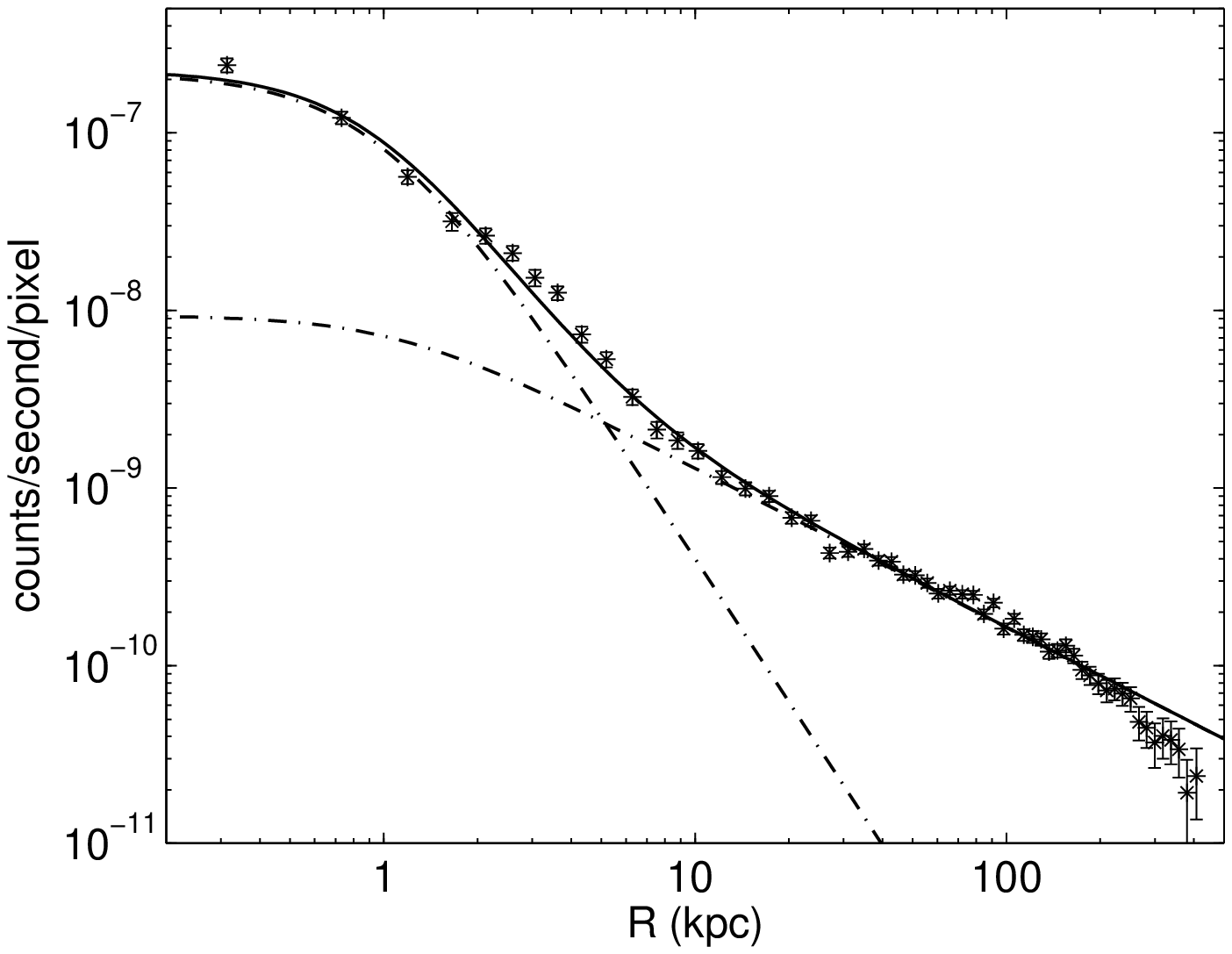}}
\end{center}
\figcaption{Background-subtracted radial profile of the 0.5--2~keV
  intragroup emission, extracted in $2\arcsec$ pixels. Solid line
  shows the best-fit double $\beta$--model, with the individual model
  components shown as dot-dashed lines. Error bars include
  uncertainties on the background level.
   \label{fig:surfbright}}
\end{figure}

Although our main goal is to study the ISM in individual group
members, it is also relevant to establish which galaxies are embedded
within detectable intragroup gas. To do so, a radial surface
brightness profile was extracted from the unsmoothed
exposure-corrected {\em Chandra} mosaic, centered on NGC\,2563 itself,
and with all other bright sources masked out. The background level was
estimated from the four outermost pointings, which are all centered at
$R=40\arcmin$--$45\arcmin$ from the X-ray peak and show consistent
full-chip count rates. The resulting background-subtracted profile,
shown in Figure~\ref{fig:surfbright}, reveals diffuse emission out to
at least $R \approx 21\arcmin$ ($\approx 400$~kpc in projection). This
is a few arcminutes further than the maximum extent determined from
{\em ROSAT} observations \citep{mulc03,osmo04}, and roughly
corresponds to our estimate of $R_{500} \approx 420$~kpc. Modeling the
profile as a sum of two $\beta$--models yields $\beta_1 =
0.61^{+0.09}_{-0.03}$, $r_{c1} = 1.0\pm 0.1$~kpc, and $\beta_2 =
0.32\pm 0.01$, $r_{c2} = 1.1^{+1.3}_{-1.0}$~kpc, for the inner and
outer component, respectively. With a reduced $\chi^2=3.0$, the fit is
poor however, in part because the profile steepens beyond that of a
$\beta$--model at large radii, as also seen in other groups and
clusters \citep{vikh06,sun09}.

The relatively high central surface brightness could potentially
compromise our ability to detect individual galaxies in the group
core. We have attempted to circumvent this issue by requiring a longer
central {\em Chandra} exposure (50~ks). In addition, the surface
brightness has dropped from its central value by three orders of
magnitude already at $R=5\arcmin \approx 100$~kpc. Apart from the
central galaxy, there are no optically luminous group members within
the central few arcmin, so our X-ray detection limits are impacted
minimally by the presence of intragroup gas. This is corroborated by
the fact that we find no systematic radial variation in limiting
$L_{\rm X,th}$ for the 54 group members covered by {\em Chandra}.  A
detailed analysis of the intragroup medium in NGC\,2563 will be the
subject of future work.

\section{Gas Stripping in NGC\,2563}\label{sec,discuss}

With our H{\sc i} and X-ray results extending well beyond the group
virial radius, we can now explore how the ISM properties of individual
galaxies vary with position across the entire group, and in particular
to what extent there is evidence for ongoing or recent ISM stripping
from the group members.

\subsection{Evidence for Ram Pressure Stripping}

The detection of X-ray tails with {\em Chandra} has provided direct
evidence for ram pressure stripping in both groups
\citep{rasm06,jelt08} and clusters \citep{sun07}, and similar
inferences have been made from the presence of H{\sc i} tails within
these systems \citep{kenn04,crow05,chun07}. In X-rays, however, the
detection rate of these features is generally very small ($<10\%$),
suggesting that strong stripping of hot ISM is either very rare or
proceeds very rapidly. No evidence for such activity in the form of
X-ray tails is seen within NGC\,2563, but this is not surprising given
the low detection rate of thermal coronae within the sample.

In H{\sc i}, we identify two tails associated with the relatively
isolated CGCG119-047 and CGCG119-051, both with H{\sc i} extending
beyond the optical disk in the direction opposite of the group center
(Figure~\ref{fig:SDSS_HI}). Neither of these objects is H{\sc i}
deficient, so any stripping activity must have recently
commenced. Their respective radial velocities relative to the group
mean of 200 and 320~km~s$^{-1}$ imply Mach numbers of ${\cal M} \ga
0.4$ and $\ga 0.6$ for an ambient gas temperature of $T\approx
1$~keV. Although both galaxies are currently beyond the radius to
which intragroup gas is detected ($R\approx 500$ and $\approx
900$~kpc), their H{\sc i} morphologies are suggestive of ongoing ram
pressure (or viscous) stripping. Multi-wavelength observations and
higher-resolution H{\sc i} data would be required to confirm this
(e.g., \citealt{murp09}). Another relatively isolated spiral, IC\,2293
at $R\sim 470$~kpc, displays no H{\sc i} tail but is highly deficient
(Def$_{\rm HI} = 0.55$), and is also a candidate for recent ram
pressure stripping.

To test for the {\em global} importance of ram pressure stripping
within the group, we next consider the radial distribution of ISM
detections. To put all galaxies on an equal footing, their thermal
X-ray luminosities and H{\sc i} masses were normalized by the galaxy
$L_K$. Excluding again the central galaxy NGC\,2563, evidence for hot
gas is seen in only five of the group members, and
Figure~\ref{fig,lxlk} shows that all of these reside relatively far
from the group center, with four out of five located beyond the radius
$R\sim 400$~kpc to which intragroup gas is detected. In contrast, no
hot ISM component is found within $R\sim 300$~kpc, where several of
the optically luminous early-types reside. This includes NGC\,2562 at
$R\sim 90$~kpc, the closest bright galaxy to the group center, with an
upper limit to $L_{\rm X,th}$ which is an order of magnitude below the
value of log\,$(L_{\rm X,th}/L_K) \simeq -4.6$ suggested by the
$L_K$--$L_{\rm X,th}$ relation for early-types in groups
\citep{jelt08}.

\begin{figure}
\begin{center}
\epsscale{1.17}
\hspace{0mm}
\plotone{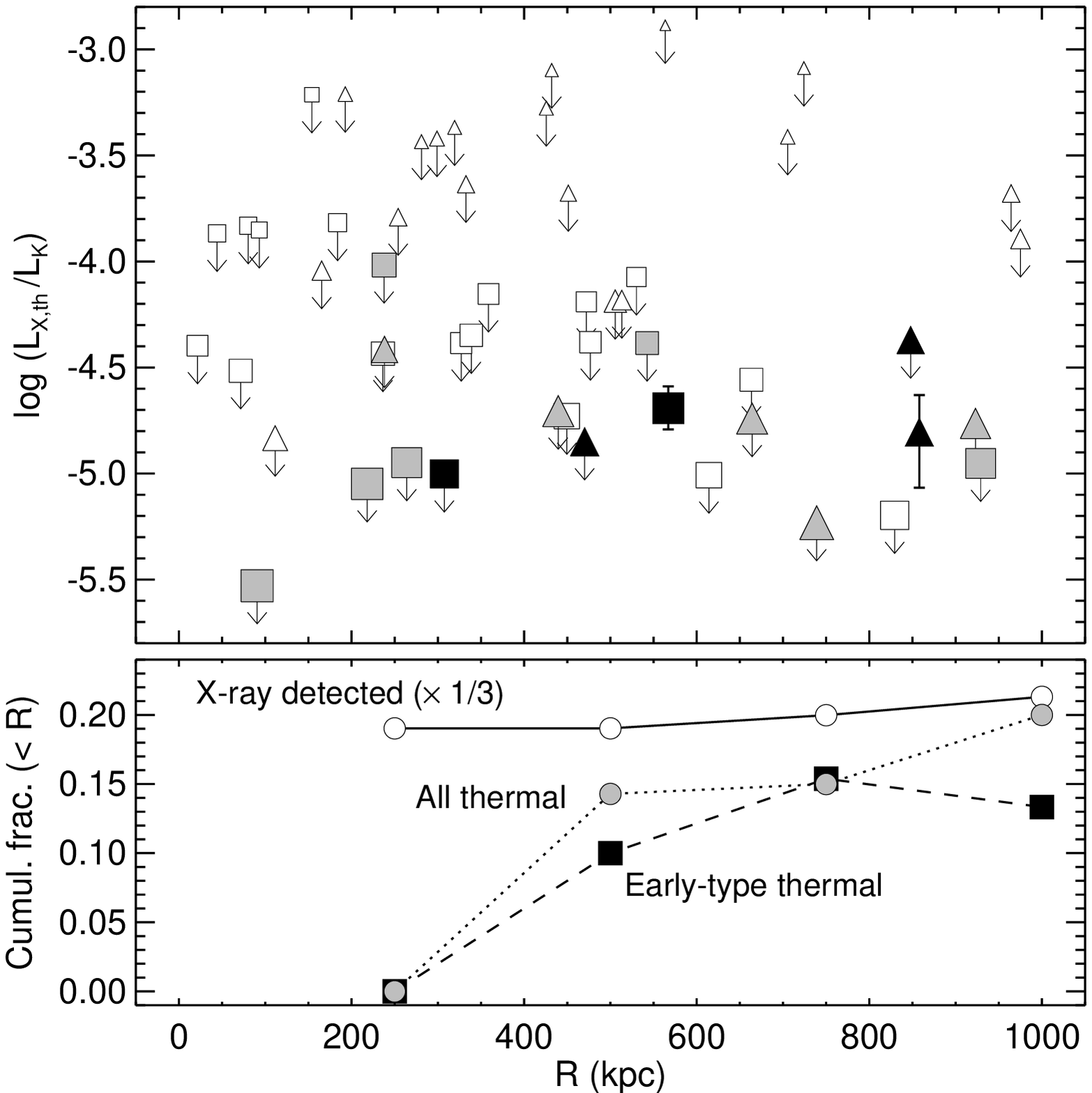}
\end{center}
\figcaption{$K$-band normalized thermal X-ray luminosity as a function
  of projected group radius $R$, excluding the central galaxy
  NGC\,2563. Symbols are as in Figure~\ref{fig,scaling}, with symbol
  sizes scaling with galaxy $L_K$. Bottom panel shows the cumulative
  fraction of optically bright ($L_K>10^{10} L_\odot$) group members
  within a given $R$ that are X-ray detected (empty symbols; scaled
  down by a factor of 3 for ease of comparison), the fraction
  containing a thermal component (grey), and that of the early-types
  doing so (black).
  \label{fig,lxlk}}
\end{figure}

To further illustrate this point, we note that, given our X-ray
detection limits (Section~\ref{sec,X}), the \citet{kim04} relation
would suggest that we should detect all early-type members brighter
than log\,$(L_K/L_\odot) \approx 10.0$, irrespective of their hot gas
content. In practice, only 8/14 ($57\pm 25$\%) are detected, even if
including NGC\,2563 itself and excluding Sa galaxies. The bottom panel
in Figure~\ref{fig,lxlk} shows the cumulative fraction of group
members covered by {\em Chandra} above this $L_K$ that is X-ray
detected, and the fraction that also contains evidence for a thermal
component. While the former remains constant with $R$, the thermal
fraction drops toward the group core. Although subject to large
Poisson errors, this result is consistent with hot ISM having been
stripped within the dense group core, while the more distant galaxies
retaining a thermal component have yet to experience peak ram pressure
during their orbit (but see also \citealt{balo00}).

In contrast, galaxies with a detectable {\em cold} ISM component are
distributed more evenly with $R$. This is illustrated in
Figure~\ref{fig,mhi_lk}(a), which compares our measured H{\sc i}
masses to those of similar galaxies in the field. A handful of
galaxies show evidence of a significant shortfall in H{\sc i} content
(i.e.\ Def$_{\rm HI}\ga 0.5$), but the values scatter broadly around
Def$_{\rm HI}=0$ and show no systematic dependence on $R$. A
Kolmogorov test was performed to test whether the inferred
deficiencies are, in fact, consistent with being drawn from a Gaussian
parent distribution centered at Def$_{\rm HI} = 0$ with some width
$\sigma$. We find that for $\sigma$ in the range 0.3--0.5, this
probability is at least $85$\%, regardless of whether the upper/lower
limits and the highly deficient CGCG119-040 are included. There is
thus no strong evidence from this that the large Sa--Sc group members
are globally H{\sc i} deficient. However, we present evidence in
Section~\ref{sec,glob} that the late-type population as a whole {\em
  is} deficient relative to the field.

\begin{figure}
\begin{center}
\epsscale{1.}
\hspace{0mm}
\plotone{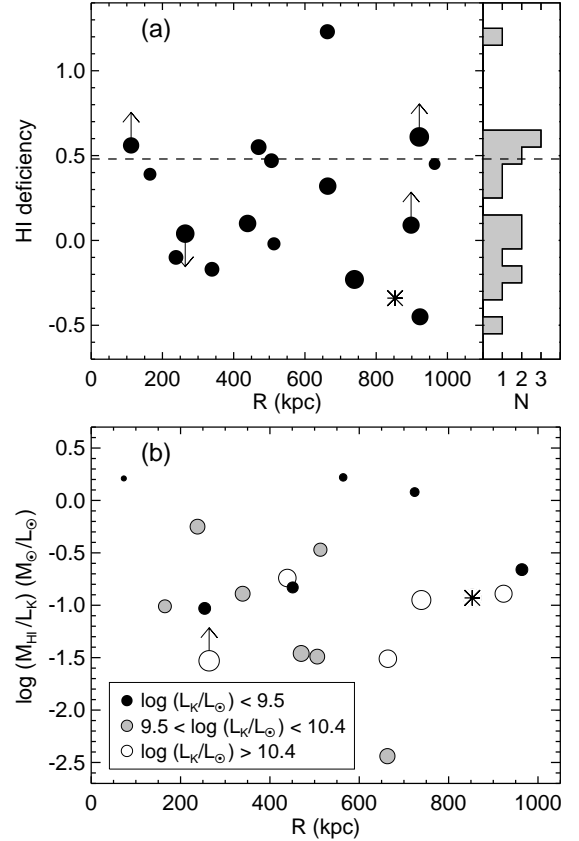}
\end{center}
\figcaption{(a) H{\sc i} deficiencies Def$_{\rm HI}$ of all large
  group spirals of types Sa through Sc, along with a histogram of
  Def$_{\rm HI}$. Galaxies above the dashed line are deficient by at
  least a factor of three. Upper limit represents UGC\,4332, which has
  incomplete H{\sc i} velocity coverage and so only a lower limit to
  its H{\sc i} mass (cf.\ Table~\ref{tab:HI}). (b) $K$--band
  normalized H{\sc i} masses for all H{\sc i} detected group members,
  color-coded according to galaxy $L_K$. Lower limit marks
  UGC\,4332. In both plots, symbol sizes scale with $L_K$, and
  asterisk represents the IC\,2338/2339 pair.
  \label{fig,mhi_lk}}
\end{figure}

The estimated H{\sc i} deficiencies depend on assumed galaxy
morphology and the somewhat uncertain comparison values for field
galaxies. A more robust measure of the relative H{\sc i} content
within the group members might be provided by $M_{\rm HI}/L_K$. The
dependence of this quantity on $R$ is shown in
Figure~\ref{fig,mhi_lk}(b). Note that at fixed $R$, lower-mass
galaxies would be more easily ram pressure stripped, but they are also
likely to be more H{\sc i}--rich to begin with. To take this into
account, the galaxies were divided into three bins of comparable
$L_K$. Figure~\ref{fig,mhi_lk}(b) confirms that even at ``fixed''
$L_K$, there is no systematic radial dependence of H{\sc i} content
within the group, and hence that ram pressure stripping of cold ISM is
unlikely to be globally important in this system.

\subsection{Evidence for Tidal Gas Stripping}

The presence of H{\sc i} tails associated with close galaxy pairs is
commonly taken as a sign of tidal encounters \citep{deme08,koop08}.
NGC\,2563 contains two galaxy pairs with clear H{\sc i} extensions
indicating such encounters. The most prominent of these is associated
with IC\,2338 and IC\,2339, a spiral pair separated by only $\sim
15$~kpc in projection and $\sim 20$~km~s$^{-1}$ in velocity. The SDSS
images show evidence for a stellar bridge between the galaxies,
supporting a tidal origin for the H{\sc i} tails. An H{\sc i}
extension is also connecting UGC\,04324 with CGCG119-040; the latter
represents the most H{\sc i} deficient group member (deficient by a
factor of $\sim 15$), with all its H{\sc i} associated with the H{\sc
  i} bridge. While no corresponding stellar feature is seen in the
relatively shallow SDSS images, the H{\sc i} morphology is strongly
suggestive of a tidal encounter.

\begin{figure*}
\begin{center}
\epsscale{0.9}
\plotone{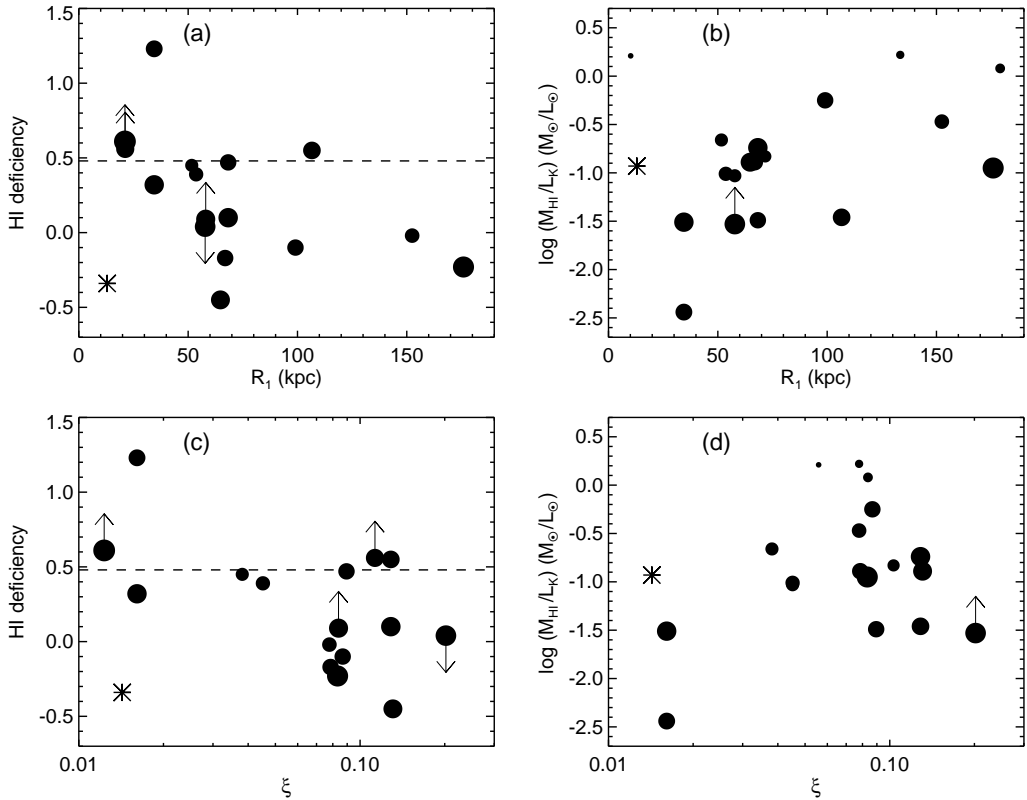}
\end{center}
\figcaption{(a) H{\sc i} deficiencies Def$_{\rm HI}$ of all large
  group spirals of types Sa through Sc as a function of projected
  distance $R_1$ to the nearest neighbor. Dashed line is as in
  Figure~\ref{fig,mhi_lk}. (b) $K$-band normalized H{\sc i} masses for
  the H{\sc i} detected objects against $R_1$. Lower limit marks UGC\,4332. 
  (c) As (a), but as a
  function of $\xi$ from equation~(\ref{eq,xi}). (d) As (b), but as a
  function of $\xi$. In all plots, symbol sizes scale with $L_K$, and
  asterisk represents the IC\,2338/2339 pair.
\label{fig:HIdef}}
\end{figure*}

To explore whether recent removal of cold ISM through tidal
interactions can generally explain the observed H{\sc i} deficiencies
within the group, we show in Figure~\ref{fig:HIdef}(a) the H{\sc i}
deficiencies as a function of projected distance $R_1$ to the nearest
neighbor. If excluding the closely interacting IC\,2338/2339 pair and
the upper/lower limits, a Kendall correlation test suggests an
anticorrelation at the $1.0\sigma$ level. Including the upper/lower
limits at their nominal values strengthens the correlation
significance to $2.2\sigma$. The correlation with $R_1$ remains
significant at the $2\sigma$ level when instead considering the
normalized H{\sc i} masses in Figure~\ref{fig:HIdef}(b). While only
indicative, this result is consistent with some H{\sc i} having been
removed in galaxy--galaxy interactions within the group.

Since strong tidal interactions require proximity of two galaxies in
both position and velocity space, a plot of H{\sc i} content against
$R_1$ may be subject to large scatter due to some galaxies being close
in projection only. For each galaxy, we therefore also consider the
minimum value of
\begin{equation}
 \xi = \sqrt{ (\Delta R/\Delta R_{\rm max})^2 + (\Delta v_r/\Delta v_{\rm
     max})^2} ,
\label{eq,xi}
\end{equation}
where $\Delta R$ and $\Delta v_r$ are the separations in projection
and radial velocity between {\em any} two group members, and $\Delta
R_{\rm max} \approx 2.1$~Mpc and $\Delta v_{\rm max} \approx
1600$~km~s$^{-1}$ are the corresponding maximum values between all
galaxies in NGC\,2563. Strong interactions would require $\xi \ll 1$,
and, if generally prominent in removing H{\sc i}, would imply a
correlation between H{\sc i} content and $\xi$ for small values of the
latter. For NGC\,2563, one might expect interactions to be important
up to at least $\xi \approx 0.1$, as this would correspond to two
galaxies at the same $v_r$ separated by $\Delta R\la 200$~kpc, or to a
physically ``overlapping'' pair with $\Delta v \la
160$~km~s$^{-1}$. As suggested by Figure~\ref{fig:HIdef}(c) and (d),
there is some evidence of such a correlation for $\xi \la 0.1$ among
the individually H{\sc i} detected galaxies. However, this is
significant at less than $2\sigma$, so we can only conclude that our
results are at least consistent with some H{\sc i} removal due to
galaxy--galaxy interactions. Repeating this analysis for the X-ray
detected group members reveals no indication of a systematic trend in
$L_{\rm X,th}/L_K$ with either $R_1$ or $\xi$. However, many of these
galaxies have X-ray luminosities only slightly above our completeness
limit, so deeper observations of a larger sample would be required to
confirm this result within groups in general.

\subsection{Global HI Properties of the Group}\label{sec,glob}

Our comprehensive VLA coverage of NGC\,2563 enables a census of the
global amount and distribution of H{\sc i} in the group, and provides
further evidence of recent H{\sc i} mass loss from the group members.
Based on measurements from the H{\sc i} Parkes All Sky Survey
(HIPASS), \citet{evol11} have quantified the typical H{\sc i} mass for
late-type galaxies (Sb and later) in the general ``field'' with
$M_\ast > 10^8 M_\odot$:
\begin{equation}
  M_{\rm HI} = M_1 \left(\frac{M_\ast}{M_2}\right)^{0.19}
  \left[1+\left(\frac{M_\ast}{M_2}\right)^{0.76}\right], 
\label{eq,HI}
\end{equation}
where $M_1=3.36\times 10^9 M_\odot$ and $M_2=3.3\times 10^{10}
M_\odot$. To compare this to results for NGC\,2563, we used estimates
of $M_\ast$ for the group members from
SDSS\footnote{http://www.mpa-garching.mpg.de/SDSS/DR7/}, available for
all but six of the 64 members; for those six, we assume a $K$--band
stellar mass-to-light ratio of 0.55, which is the average for the
members with both $L_K$ and $M_\ast$ available.
Equation~(\ref{eq,HI}) would then suggest a total H{\sc i} mass of the
relevant group members of $6.7 \pm 2.0 \times 10^{10} M_\odot$,
whereas the value inferred from Table~\ref{tab:HI} is $M_{\rm HI} <
4.4\times 10^{10} M_\odot$, when including the average $3\sigma$ upper
limit of $2\times 10^8M_\odot$ for the undetected members. This
suggests that globally, the $M_\ast > 10^8 M_\sun$ late-types in the
group have suffered mild H{\sc i} mass loss with respect to similar
field galaxies. Compared to estimates of Def$_{\rm HI}$, this result
may represent a more robust estimate of global H{\sc i} deficiency
within the group, as it includes also smaller galaxies, depends less
sensitively on the inferred morphology and optical size of individual
group members, and takes the dispersion seen for field galaxies into
account.

We also derive a total stellar mass in the group of $M_\ast \approx
1.1\times 10^{12} M_\odot$ for all the 64 spectroscopically identified
members, and a corresponding total range of $M_{\rm HI} = (5.7-
6.5)\times 10^{10} M_\odot$. This implies $M_{\rm HI}/M_\ast
=0.054$--0.062 for the full group (down to our 98\% spectroscopic
completeness limit of $M_r = -17$), under the plausible assumption
that our VLA observations are not missing any large-scale diffuse
H{\sc i}. Future comparison of this result for an X-ray bright group
to other groups and clusters could provide further information on the
nature of gas removal in these environments.

On the largest scales, the distribution of H{\sc i}--detected galaxies
in Figure~\ref{fig,hi_mosaic} reveals a peculiar lopsidedness, with 16
of the 20 detections occurring on the western side of the
group. This is not simply due to asymmetric spatial VLA coverage, as
demonstrated in Figure~\ref{fig:HIchannel}, nor to a larger fraction
of H{\sc i} deficient galaxies in the east. The wide velocity coverage
of our VLA data ($\Delta v_r \approx \pm 3\sigma_{\rm biwt}$) further
renders it unlikely that a significant fraction of potential group
members have been missed on the eastern side of the system. The
lopsided H{\sc i} distribution thus reflects a real asymmetry in the
distribution of late-type members, with nearly all the large spirals
located in the western half of the group.

The origin of this asymmetry remains unclear. One possibility is that
the western spirals are part of an infalling subgroup, but echoing an
earlier conclusion by \citet{zabl98b}, we find no evidence for
kinematic substructure in the group using the technique pioneered by
\citet{dres88}. Another possible scenario is that galaxy morphologies
have been modified by strong interactions with a similarly asymmetric
intragroup medium. This can be ruled out, however, because the
intragroup gas distribution in NGC\,2563 appears fairly symmetric on
large scales and, if anything, seems more strongly extended to the
west (see Figure~\ref{fig:mosaic} and \citealt{mulc98}). Finally,
exploiting the extensive SDSS coverage in this field, we have also
examined the overall galaxy distribution within a projected distance
of $R=5$~Mpc of the group center, and find no evidence for a general
large-scale enhancement in galaxy density on the western side of the
group. Further studies of other groups on very large scales may help
establish how common such morphological segregations are in these
systems.

\section{Summary and Conclusions}\label{sec,summary}

Our extensive {\em Chandra} and VLA coverage of the NGC\,2563 galaxy
group has enabled us to probe both the hot and cold ISM within a
nearby group out to $\sim 1.4$ times the estimated virial radius. The
main aim has been to characterize the evidence for galactic gas
removal in this system, and understand the mechanisms involved.
Although the limited number of galaxies preclude strong conclusions
based on a single group, our results suggest that both hot and cold
gas have been stripped from some of the group members, and that both
ram pressure stripping and gas removal via galaxy--galaxy interactions
are all occurring simultaneously within this one system. This is based
on the following evidence:

(i) The thermal X-ray deficiency of optically luminous early-type
galaxies in the group core suggests recent ram pressure stripping of
their hot gas. The central group galaxy aside, no hot ISM is detected
within $R\sim 300$~kpc from the group core, whereas four out of the
five non-central galaxies with such a component reside beyond the
radius to which intragroup gas is detected. The thermal luminosity of
the bright early-type closest to the group core (NGC\,2562) is at
least an order of magnitude below that expected for typical group
galaxies of the relevant $L_K$ \citep{jelt08}.

(ii) Comparison of the H{\sc i} content of the late-type group members
to that of similar galaxies in the field suggests that the former are,
on average, mildly deficient in H{\sc i}. This points to one or more
mechanisms removing cold ISM within the group.

(iii) Ram pressure (or viscous) stripping of cold gas is suspected in
a few cases. The H{\sc i} data reveal one relatively isolated spiral
(IC\,2293) which is H{\sc i} deficient by more than a factor of
three. Another two such spirals show H{\sc i} morphologies suggestive
of an ongoing ram pressure interaction, with H{\sc i} tails pointing
away from the group core (CGCG\,119-047 and 119-051 in
Figure~\ref{fig:SDSS_HI}). However, these are not (yet) H{\sc i}
deficient by the usual definition, indicating that significant
interactions with the intragroup medium may have only recently
commenced.

(iv) Ongoing galaxy--galaxy interactions removing H{\sc i} are also
strongly suggested in at least two cases. The two most prominent H{\sc
  i} tails/extensions in the group occur within close galaxy pairs,
and the most H{\sc i} deficient group member (CGCG\,119-040, deficient
by a factor of $\sim 15$) is a member of one of these.  Suggestive
evidence is further seen for galaxies with close neighbors in
position--velocity space to show relatively low H{\sc i} content,
consistent with tidal stripping of H{\sc i}.

The inference that ram pressure stripping of hot galactic gas may have
occurred in the central group regions would be in line with other {\em
  Chandra} studies of group and cluster galaxies \citep{rasm06,sun07},
and with simulations which suggest that such stripping can be
efficient even in small galaxy groups \citep{kawa08}. However, it is
worth noting that a large fraction of early-type galaxies generally do
retain halos even in X-ray bright groups, including some in NGC\,2563,
and that such halos are not generally underluminous compared to those
of galaxies in the field \citep{jelt08,mulc10}. Despite the detection
of a hot intragroup medium within NGC\,2563 out to $R\approx R_{500}$,
there is also no global evidence for ram pressure stripping of the
{\em cold} ISM within the group. Specifically, no radial trends are
seen in the stellar mass--normalized H{\sc i} content among the 20
H{\sc i} detected group members.

The indicative result that ram pressure may affect only the hot ISM in
typical group galaxies is consistent with existing numerical studies
\citep{kawa08,rasm08}. Galaxy--galaxy interactions remain the main
candidate for removing cold ISM within NGC\,2563 and explaining the
global H{\sc i} deficiency inferred for the late-type group members.
Other studies have also implied that tidal encounters have an
important impact on the H{\sc i} properties of galaxies in groups
\citep{kern08,rasm08}. In addition, such encounters may work to
enhance the susceptibility of a galaxy to ram pressure stripping, by
perturbing the distribution of the cold gas
\citep{davi97,maye06}. Nevertheless, despite the detection of two new
group members from our optically blind H{\sc i} search, we find no
evidence within the NGC\,2563 group for isolated, optically dark H{\sc
  i} clouds that might represent previously removed material.

Given the small number of luminous galaxies in a single group, similar
studies of a larger group sample will be required to better understand
how the group environment impacts galaxy evolution. For example, we
note that the inferred H{\sc i}-to-stellar mass ratio of $\approx
0.06$ inferred for this X-ray bright group may provide a useful
benchmark for the H{\sc i} content of dynamically evolved
groups. Future comparison of this result to those of X-ray faint
systems, richer clusters, and groups at higher redshift (of which the
upcoming Square Kilometre Array should detect many thousands) could
further improve our understanding of the nature and epoch of gas
removal in dense environments. Such studies would also help to address
the commonality of the highly lopsided distribution of H{\sc i} seen
in NGC\,2563, which reflects a puzzling galaxy morphological
segregation within the group.

\acknowledgments

We are grateful to the referee for a careful and constructive report
which significantly improved the presentation of our results. We thank
Christy Tremonti for providing the emission-line ratios for the SDSS
spectra. This research has made use of the NASA/IPAC Extragalactic
Database (NED).  JR acknowledges support by the Carlsberg
Foundation. XB thanks Prof.\ Zhang, S.~Nan, and R.~Fengyun from
Tsinghua Center for Astrophysics (THCA) for discussions on {\em
  Chandra} analysis. Support for this work was provided by the
National Science Foundation under grant number 0607643 to Columbia
University, and by the National Aeronautics and Space Administration
through Chandra Award Number G07-8134X issued by the Chandra X-ray
Observatory Center, which is operated by the Smithsonian Astrophysical
Observatory for and on behalf of the National Aeronautics Space
Administration under contract NAS8-03060.

\end{document}